\documentstyle[preprint,aps,eqsecnum,epsfig]{revtex}
\tighten
\begin{document}
\preprint{PITT-97-333; LA-UR-98-??; LPTHE-98-05,PE-97-??}
\draft
\title{\bf EVOLUTION OF INHOMOGENEOUS CONDENSATES:\\
SELF-CONSISTENT VARIATIONAL APPROACH}
\author{\bf D. Boyanovsky$^{(a)}$, F. Cooper$^{(b)}$,
H. J. de Vega$^{(c)}$, P. Sodano$^{(d)}$}
\address
{(a) Department of Physics and Astronomy, University of
Pittsburgh, Pittsburgh  PA. 15260, U.S.A \\
(b) Theoretical Division, Los Alamos National Laboratory,
 Los Alamos, NM 87545, U.S.A. \\
 (c) LPTHE, Universit\'e Pierre et Marie Curie (Paris VI) et Denis Diderot
(Paris VII), Tour 16, 1er. \'etage, 4, Place Jussieu, 75252 Paris, Cedex 05,
France \\
(d)Dipartimento di Fisica and Sezione INFN, Universit\`a di
Perugia, 06100 Perugia, Italia}
\date{February 1998}
\maketitle
\begin{abstract}
We establish a self-consistent variational framework that allows us to study
numerically the non-equilibrium evolution of non-perturbative inhomogeneous
field configurations including quantum backreaction effects. After discussing 
the practical merits and disadvantages of different approaches we provide a
closed set of local and renormalizable
update equations that determine the dynamical evolution of inhomogeneous
condensates and can be implemented numerically. These  incorporate
self-consistently the backreaction of quantum fluctuations and particle
production. This program requires the solution of a self-consistent
inhomogeneous problem to provide initial Cauchy data for the
inhomogeneous condensates and Green's functions. We provide a simple solvable
ansatz for such an initial value problem for
the Sine Gordon and $\phi^4$ quantum field theories in one spatial
dimension. We compare exact known results of the Sine Gordon model
to this simple ansatz. We also study the linear sigma
model in the large $N$ limit in three spatial dimensions as a microscopic
model for pion production in ultrarelativistic collisions.
We provide a solvable self-consistent ansatz for the
initial value problem with  cylindrical symmetry.
For this case we also obtain a
closed set of local and renormalized update equations that can be
numerically implemented. A novel phenomenon of spinodal instabilities and
pion production arises as a result of a Klein paradox for large
amplitude inhomogeneous condensate configurations.

\end{abstract}

\section{Introduction}
The dynamical evolution of spatially inhomogeneous semiclassical field
configurations along with quantum fluctuations that dress them play an
important role in many relevant problems in high energy and condensed matter
physics. Such situations are ubiquitous in the description of semiclassical
processes such as  baryogenesis via sphaleron
decay\cite{kuzmin,rebbi}, bubble and droplet nucleation during supercooled
phase transitions as it could be the case    in QCD\cite{meyer,muller}, the
formation of topological defects\cite{kibble,vilenkin} and the possibility of
 formation of disoriented chiral
condensates\cite{anselm,blaizot,bjorken,kowalski}.  Localized high energy and large 
amplitude field configurations are of  particular relevance for modelling
the dynamics of relativistic heavy ion collisions as was proposed for
example by\cite{humuller} in terms of collisions of Yang-Mills wave-packets.
In this case an approximation to the
dynamics of high initial multiplicity states would be to provide an
initial value problem with a particular initial condition on the field
at some proper time surface. By proposing a large amplitude field
configuration localized in space in the form of a semiclassical coherent
state, the features associated with an initial state of large energy density
and high multiplicity are captured. For example such field could
be a strong color field or electric field whose relaxation will result
in high multiplicity particle production\cite{humuller}.

The dynamics of coherent structures and semiclassical field
configurations  is also  of theoretical and experimental relevance in
condensed matter physics within the context of transport phenomena mediated
by soliton excitations in quasi-one dimensional
systems\cite{schrieffer,yu,gruner}.

Although there have been several studies of real-time dynamics of classical
field configurations within the context of sphaleron decay\cite{rebbi},
Non-Abelian wave-packet collisions\cite{humuller} and coherent structures in
scalar field theories\cite{gleiser} including classical\cite{rajagopal,gavin,asakawa}
and semiclassical\cite{randrup} studies of disoriented chiral condensates,
there have been very few and limited attempts to incorporate consistently the 
effects  of quantum fluctuations upon the dynamics of semiclassical
inhomogeneous configurations and mainly in one spatial
dimension\cite{vautherin}. Whereas the dynamics of inhomogeneous
field configurations had been studied in the small amplitude
regime\cite{inhomo}, the important regime of large amplitude and
non-perturbative field configurations is still not understood.

Recently there has been tremendous progress in studying the non-equilibrium
dynamics of high energy and high multiplicity quantum states including
self-consistently quantum fluctuations in the case of
spatial or spatial rapidity homogeneity to study particle production in
strong fields, dynamics of phase transitions and relaxation and the dynamics
of inflation\cite{primo}-\cite{tsunami}.  In particular these methods had been 
used to study the formation of disoriented chiral
condensates, and pion production and relaxation during the chiral phase
transition in low-energy phenomenological models\cite{dcclanl,boydcc} and more
recently initial quantum states of large multiplicity distributions with
localized momenta\cite{tsunami}.

These studies rely on an implementation of variational and large
$ N $\cite{fred}-\cite{largeN} approximations generalized to strongly out of
equilibrium situations. These approximations \cite{jackiw}-\cite{camelia}   
are specially suited for studying non-perturbative
phenomena and in particular the large$ N $approximation is amenable to
consistent improvement\cite{losalamos,largeN}.

Since these methods allow one to study the full quantum evolution in a
non-perturbative manner they are specially tailored to studying the
non-equilibrium evolution of initial inhomogeneous quantum states of large  
energy including quantum back-reaction effects.

This is specially important in the modelling of relativistic heavy ion
collisions in a manner that allows one to include the effects of quantum
backreaction and  non-perturbative particle production.
A useful description of the dynamical evolution in relativistic heavy ion
collisions is based on a hydrodynamic picture in terms of a fluid that is
described in local thermodynamic equilibrium by a local energy momentum
tensor and an equation of state\cite{fred2,bjorken2}. Although such a
description is not only physically
compelling but also experimentally consistent at the
present energies\cite{meyer,books},  it is at best phenomenological and not  
justified from a microscopic theory. The hydrodynamic description
relies on the assumption  that as a result of the
collision a large amount of energy is deposited in a small region of
space-time. Boost invariant hydrodynamics\cite{fred2,bjorken2} should emerge
when the energy density is much
larger than the typical hadronic masses, and is a result of hydrodynamic
scaling when the resulting pancake width goes to zero with center of mass  
energy.  In this scaling limit the relevant degrees of freedom for describing  
the expansion are the fluid proper time and fluid
rapidity.

Although trying to study the emergence of boost invariant hydrodynamics
within the framework of QCD is at this stage an impossible task, the features 
of hydrodynamical evolution and particle production are rather robust and
should be captured by a microscopic albeit semi-phenomenological theory.

 In this article
we begin our study of the non-equilibrium dynamics of inhomogeneous field
configurations of large amplitude including the back-reaction of quantum
fluctuations  self-consistently and non-perturbatively. The focus
and goal of this program is to study the relaxation of a strongly out of
equilibrium inhomogeneous initial state,  the  description of particle
production and the emergence of a hydrodynamical description from an
{\em ab initio} full quantum evolution in a microscopic, phenomenologically
inspired quantum field theory.

This article is devoted to setting up the variational self-consistent problem 
and to studying the subtle but important issues of renormalization. The
implementation of variational or large $ N $ approximations to study the
dynamics requires:  i) to determine the Cauchy data for the inhomogeneous
expectation value of the field as well as the
relevant Green's functions that account for backreaction. Since both the
variational and the large $ N $ approximations lead to a self-consistent
condition, the Cauchy data must be determined self-consistently. The
solution to this self-consistent initial value problem for inhomogeneous
field configurations is typically an extremely difficult and numerically
intensive task.
 ii) setting up a closed set of {\em local} renormalized update equations
that can be numerically implemented in current computers. Both the locality
as well as renormalizability
aspects are extremely important: current computational capabilities
cannot handle non-local update equations because they are memory
intensive. The numerical evolution will be implemented on a space-time
mesh and the results should be insensitive to the mesh size, this
requires renormalized evolution equations, which is also important to
improve numerical stability. This issue is clearly very important for
the quantum backreaction because the Green's functions that are
necessary have short distance singularities that must be renormalized.

iii) Finally the numerical implementation of the update equations
obtained in ii) starting from the initial data provided by the first
step.

In this work we seek to provide the first two steps to implement this
program, leaving the numerical implementation to a forthcoming article.

The coupled evolution equations that we find more suitable for a numerical 
approach in the self-consistent Hartree and large N approximations are summarized
below in eqns. (\ref{classi}-\ref{kdot2}). 

We focus on studying the initial value problem and the update
equations first in relevant cases in $1+1$ dimensions not only because
it is a simpler setting but also because it is a relevant problem not
only in quantum field theory but also in condensed matter physics in
the case
of quasi-one dimensional systems. By treating these examples in one
spatial dimension we determine a strategy to set up the self-consistent
initial value problem that can be implemented for particular three
dimensional geometries  relevant to the description of `colliding
pancakes' in heavy ion collisions. This strategy is based on mapping
the initial value problem (in terms of the variational parameters) onto a
Schr\"odinger-like problem whose spectrum of eigenvectors and
eigenvalues is exactly known.

We then move on to the linear sigma
model description of pion physics and implement a large $ N $ self-consistent
approximation and the strategy developed in the one dimensional case to
set up the Cauchy data. In this case we find some novel phenomena
associated with spinodal instabilities for large amplitude field
configurations. Finally we address the issue of renormalization and
obtain the (semi) local update equations.

The article is organized as follows: in section II we introduce the
different formulations to implement the variational and Hartree approximations, discussing the advantages and disadvantages of each of the methods and determine the method that we find the most suitable for numerical implementation at this stage. 

 Section III is devoted to a
description of
the initial value problem. Section IV presents a detailed analysis
of the self-consistent initial value problem for the cases of the
Sine-Gordon and $\phi^4$ theories in $1+1$ space-time
dimensions. After analyzing the vacuum structure in the variational
approximation we propose simple  self-consistent Ansatzse
for the  Cauchy data for both
topological (kink) and pulse-like (lumps) field configurations by
establishing a correspondence between the initial value problem in
terms
of the variational parameters and the eigenvalue problem of a
Schr\"odinger operator whose spectrum is exactly known. The
renormalization aspects are studied in detail. These examples
are relevant for studying the dynamics of solitons dressed by the
quantum fluctuations in $1+1$ dimensions and to study the relaxation of
large amplitude field configurations via meson production.

In section V we study the linear sigma model description of low energy pion  
phenomenology by implementing a large $ N $ approximation in $3+1$
dimensions. We find that the leading order in the large N is similar to
a Hartree factorization which allows us to use the formulation presented
in section II. 

 We use the results of the $1+1$ dimensional case to set up
self-consistent
Cauchy data for an initial inhomogeneous configuration with cylindrical
symmetry that could model the initial state after the collision of
Lorentz contracted `pancakes'. The subtle renormalization aspects are
discussed in detail and we find a phenomenon akin to the Klein
paradox that results in spinodal instabilities for large amplitude
inhomogeneous configurations.

In section VI the dynamical evolution equations are summarized in terms
of fully renormalized quantities and the initial data on all of the
Green's functions that are needed for the update is
determined. Section VII proposes a numerical strategy for an
implementation of the dynamics.
We summarize our strategy for the present and future work in the conclusions.


\section{Non-Equilibrium Dynamics: Methods}
There are several alternative  ways to establish a variational  
self-consistent formulation of the dynamical evolution, and although all of  
them are equivalent,
some particular implementation may  present some practical advantages over others.

Thus we begin the program of setting up the variational self-consistent  
formulation by first studying the different implementations, discussing their
merits and drawbacks, and choosing a particular formulation that would lead  
to an economical numerical implementation.

\subsection{ Time Dependent Variational Principle}
The method for obtaining a time dependent {\em variational approximation}
to the non-equilibrium dynamics in  quantum field theory is to start
with Dirac's variational principle for obtaining the functional
Schr\"odinger equation.

This variational approach is implemented by proposing Gaussian wave
functionals\cite{losalamos,jackiw}-\cite{camelia}, whose parameters
are determined self-consistently. This approach leads to the time
dependent Hartree approximation, and becomes exact in the case of a
scalar field in the vector representation of the $O(N)$  group in the
large $ N $ limit\cite{losalamos,largeN}.

 Dirac's variational principle proposes to extremize  the effective action
\begin{equation}
\Gamma = \int dt < \Psi | i {\partial
\over \partial t} -H| \Psi > \; \; ; \; \; \langle \Psi | \Psi \rangle =1
\end{equation}
and leads directly to the Schr\"odinger equation:
\begin{equation}
 \delta\Gamma = 0 \rightarrow
\{ i {\partial \over \partial t} - H \} |\Psi > = 0
\end{equation}
In the Schr\"odinger ($\varphi$) representation, the wave-functionals
are given by
\[ \Psi[\varphi,t] = < \varphi |\Psi > \]
and the functional Schr\"odinger equation becomes
\begin{equation}
i\frac{\partial}{\partial t} \Psi[\varphi,t]=
 \int d^d x \left\{ -{1 \over 2} \frac{\delta^2}{\delta \varphi(x)^2}+
 {1 \over 2} \nabla
\varphi(x) \nabla \varphi(x) + V[\varphi] \right\} \Psi[\varphi,t] \label{ham}
\end{equation}
Where we have specified $d$-spatial dimensions. In later sections we will
address $d=1$ and $d=3$ separately in specific examples.

In this representation the variational approach is implemented by
proposing  a Gaussian or Hartree trial wave
functional\cite{losalamos,jackiw}-\cite{camelia}:
\begin{eqnarray}
\Psi_v[\varphi,t]& = &  N(t)\exp \left\{ -\int d^dx \int d^d y \left\{
(\varphi(\vec x)-  \phi_c(\vec x,t)){\cal K}(\vec x,\vec y;t)(\varphi(\vec y)-
 \phi_c (\vec y,t))\right\}+ \right. \nonumber \\
&  & \left.
 i \int d^d x \; \pi_c(x,t) (\varphi(x)-  \phi_c(x,t))\right\}
\label{eq:trial}\\
{\cal K}(\vec x,\vec y;t) & = &  G^{-1}(x,y,t)/ 4-i\Sigma(x,y,t) \label{kernel}
\end{eqnarray}
where in (\ref{kernel}) we have written the kernel
${\cal K}(\vec x,\vec y;t)$ explicitly
in terms of its real $G$ and imaginary $\Sigma$ parts. The real quantities
 $\varphi_c \; ; G \; ; \Sigma \; ; \pi_c$ are variational  and
have the following meaning:

\begin{eqnarray}
 && \phi_c(x,t)= < \Psi_{v}|\varphi (x) | \Psi_{v} >\; ; \; \;   \pi_c(x,t) =
<\Psi_{v}|-i \delta/\delta\varphi(x)| \Psi_{v} >\cr \cr
 && G(x,y,t) = < \Psi_{v}|
\varphi(x)\varphi(y)| \Psi_{v} > -  \phi_c(x,t)\;  \phi_c(y,t) \nonumber \\
&&< \Psi_{v}|
\varphi(x,t) ~ {1 \over i}{\delta \over \delta\varphi(y,t )} | \Psi_{v} > -   
\phi_c(x,t)\;
\pi_c(y,t) = 2\int d^dz ~G(x,z;t) \Sigma(z ,y;t)  + {i \over 2} \delta^3 (x-y)
\nonumber \\
\end{eqnarray}
The subindex $ _v $ stands for variational.

In terms of these parameters  the effective action for the trial wave
functional becomes
\begin{eqnarray}
\Gamma( \phi_c,  \pi_c,G,\Sigma) &&=\int dt < \Psi_{v}| i
\partial/ \partial t -H| \Psi_{v}> \nonumber \\
&&= \int dt d^dx\; \left\{\pi_c(x,t)\frac{\partial \phi_c(x,t)}{\partial t}
+\int dt d^dx d^dy
\Sigma(x,y;t )\frac{\partial G(x,y; t)}{\partial t} \right\}\nonumber \\
&&-\int dt < H >
\end{eqnarray}
where
\begin{eqnarray}
< H >&& = \int d^dx \{{1 \over 2} \pi^{2} + 2 [\Sigma G\Sigma](x,x) +
{1 \over 8}
G^{-1}(x,x) + {1 \over 2} (\nabla \varphi)^{2} \nonumber \\
&& + {1 \over 2}
\lim_{x \rightarrow y}  \nabla_x \nabla_y G(x,y)  + <V[\varphi(x)]> \}\; .
\label{effham}
\end{eqnarray}

\noindent $< H >$ is a constant  and is a first integral of the motion.
The equations that follow by varying the effective action with respect to
the variational parameters are:
\begin{eqnarray}
\dot{\pi}_c(x,t)&&= \nabla^{2}\phi_c(x,t)- {\partial <V > \over \partial
\phi_c }(\vec x,t);\nonumber \\ \dot{\phi}_c (x,t) &&= \pi_c(x,t)  \nonumber \\
\dot{G}(x,y; t)&&= 2\int d^dz[\Sigma (x,z)G(z,y)+G(x,z)\Sigma (z,y)]
\nonumber \\
\dot{\Sigma}(x,y;t)&&=  \int d^dz[-2 \Sigma (x,z)\Sigma (z,y) +
{1 \over 8} G^{-1}
(x,z)G^{-1} (z,y) ]\nonumber \\
&& +  [{1 \over 2}\nabla^{2}_{x}
- {\partial < V > \over \partial G}(\vec x,t)] \delta(x-y)   \label{tdhf}
\end{eqnarray}

These equations form a closed system and are the Time Dependent Hartree
(Fock) equations (TDHF). They have the drawback that they
are non-local and involve not only the Green's function $G(x,y;t)$ but
also its inverse, furthermore the renormalizability aspects of the dynamical
equations are not transparent in this formulation.
 A numerical implementation of an initial value problem
must provide Cauchy data for the variational parameters and implement
a numerical update procedure that yields finite quantities, i.e. is free of
short distance divergences.

There is an {\em equilibrium} by product of this method in the form of the
variational
effective potential for the case of static  homogeneous configurations which
is given by\cite{jackiw,camelia}
\begin{equation}
V_{eff}(\phi_c) = \frac{1}{\Omega} < \Psi|H|\Psi> \; \; ; \; \;
<\Psi|\varphi|\Psi> = \phi_c \; \; ; \; \; <\Psi|\Psi> =1 \label{veff1}
\end{equation}

\noindent with $\Omega$ being the quantization volume and $<H>$ is given in  
equation (\ref{effham}). This effective potential is useful  to establish the  
phase
structure of the particular theory in this Gaussian variational approximation. 

\subsection{ Method of Equal- time Green's Functions }

The update equations for the kernels that appear
in the Time Dependent Hartree variational wave function are non-local as
displayed by the set of equations
(\ref{tdhf}).
We can trade this non-locality by increasing the number of equations. This is 
achieved
 by considering the first order update equations for the equal time
Green's functions in terms of the Heisenberg field operator $\varphi$ and its 
canonically conjugate momentum $\pi=\dot{\varphi}$. Consider the evolution
equation of the following quantities,

\begin{eqnarray}
 && \phi_c(\vec x,t)= < \varphi (\vec x,t) > \;\;  ; \;  \; \pi_c(\vec x,t)=
< \dot{\varphi}(\vec x,t) > \nonumber \\
&& G(\vec x,\vec y,t) = < \varphi(\vec x,t)\varphi(\vec y,t) > -
\phi_c(\vec x,t)  \phi_c(\vec y,t) \nonumber \\
 && D (\vec x,\vec y,t) = < \varphi(\vec x,t) \pi(\vec y,t) >-
\phi_c(\vec x,t)\pi_c(\vec y,t)\nonumber \\
&& \tilde{D}(\vec x,\vec y,t)= < \pi(\vec x,t) \varphi(\vec y,t )  > -
\phi_c(\vec y,t)
\pi_c(\vec x,t) = D (\vec y,\vec x,t)-i\delta^d(\vec x-\vec y) \nonumber \\
&&  K (\vec x,\vec y,t) = < \dot \varphi(\vec x,t) \dot \varphi(\vec y,t) > - 
\pi_c(\vec x,t)  \pi_c(\vec y,t).  \label{kernels}
\end{eqnarray}

We can relate $D$ $\tilde{D}$ and $K$ to $G$ and $\Sigma$ as follows:
\begin{eqnarray}
D(x,y;t)+\tilde{D}(y,x;t) && =  4 \int d^dz  G(x,z;t) \Sigma(z,y;t) \nonumber \\
\tilde{D}(x,y;t)+D(y,x,t)&& =  4 \int d^dz  \Sigma (x,z;t)  G (z,y;t) \nonumber \\
\end{eqnarray}
\begin{equation}
K(x,y;t) = {1 \over 4} G^{-1} (x,y;t) + 4 \int d^dz_1 d^dz_2 \Sigma(x,z_1;t)
G(z_1,z_2;t) \Sigma(z_2,y;t) \label{eq:Kconst}
\end{equation}
This last equation, eq. (\ref{eq:Kconst}) shows that $K$ is not an  
independent quantity.

\noindent All of the expectation values are with respect to the {\em
initial} Gaussian trial wave-functional or density matrix.

The time
derivatives of operators are obtained from the
Heisenberg equations with the Hamiltonian

\begin{equation}
H = \int d^d x  \; \{ {1 \over 2} \pi(x)^2 + {1 \over 2} \nabla \varphi \cdot  
\nabla \varphi+ {\mu^2 \over 2}  \varphi^2 + V[\varphi(x)] \}
\end{equation}
and the canonical commutation relations between the field $\varphi$ and
its canonical conjugate
momentum $\pi$.

These lead to the Heisenberg operator equations of motion:
\begin{eqnarray}
\pi(x,t)&& = \dot{\varphi}(x,t) \nonumber \\
\dot{\pi}(x,t)&& = \nabla^2 \varphi(x,t) - \partial V /\partial \varphi(x,t)
\end{eqnarray}

 The  evolution equation for $ \phi_c$ is as given before
\begin{equation}
 \partial^2  \phi_c(x,t) +  < {\partial V \over \partial \varphi(x,t)} > =0 \label{ficeq}
\end{equation}
and we obtain the following  coupled equations for the kernels
$G,D,\tilde{D}$ and $K$:

\begin{equation}
\dot{G}(\vec x,\vec y;t) = D(\vec x,\vec y;t) + \tilde{D} (\vec x,\vec
y;t)  \label{eq:gdot}
\end{equation}
\begin{equation}
\dot{D}(\vec x,\vec y;t) =  K(\vec x,\vec y;t) +  \nabla_y^2
G(\vec x,\vec y;t)-
< \{\varphi(\vec x,t)
-\phi_c(\vec x,t) \} {\partial V \over \partial \varphi(\vec y,t)} >
\label{eq:ddot}
\end{equation}

\begin{eqnarray}\label{eq:kdot}
\dot K(\vec x,\vec y;t) &&=  \nabla_x^2 D(\vec x,\vec y;t) + \nabla_y^2
\tilde{D}(\vec x,\vec y;t)
\nonumber \\
 &&-   < \{ \dot{\varphi}(x,t)- \dot{\phi}_c(x,t) \}
{\partial V \over \partial \varphi(y,t)} > + < {\partial V \over \partial
\varphi(x,t)}\{ \dot{\varphi}(y,t)- \dot{\phi}_c(y,t) \}  >   \nonumber \\
 && + < {\partial V \over \partial \varphi(y,t)}\{ \dot{\varphi}(x,t)-
\dot{\phi}_c(x,t) \}
 > + < \{ \dot{\varphi}(y,t)- \dot{\phi}_c(y,t) \}{\partial V \over \partial
\varphi(x,t)}  > \label{kdot}  
\end{eqnarray}

The set of eqns. (\ref{ficeq}-\ref{kdot}) is equivalent to the TDHF equations (\ref{tdhf}). 

If we evaluate all the equations of motion  values using a Gaussian Density
Matrix, the equations for $\phi_c$, $G$, $D$ and $K$
close, which is the usual truncation of the Dyson equations in a mean field
approximation.

We would like to show that the  equation for $K$ guarantees in the update
that the  quantity
\begin{equation}
 N(x,y,t) \equiv K(x,y;t)- {1 \over 4} G^{-1} (x,y;t) - 4 \int d^dz_1 d^dz_2  
\Sigma(x,z_1;t) G(z_1,z_2;t) \Sigma(z_2,y;t) \label{eq:N}
\end{equation}
is conserved in time.
In the spatially homogeneous case we showed \cite{noneq} that $N(x,y,t)$ was  
related to the initial number of particles present at $t=0$. If we are in a  
pure state such as one described by  a Gaussian wave function,  $ < a^{\dag} a  
> =0$ at $t=0$ as will be discussed below. A nonzero value of $N$
can be obtained if we start with a Gaussian density matrix as discussed in
\cite{PRD96}.

 First one can show explicitly, for the field theories discussed in this
article,  using the equations of motion
for $G$ and $\Sigma$ that for $K$ defined by eq.(\ref{eq:Kconst}) $K$ obeys eq.
(\ref{kdot}) so that the equation for $K$ is automatically satisfied if
we solve the TDHF equations (\ref{tdhf}).  Conversely, we can rewrite the equation
for $N$  (\ref{eq:N}) in terms of $D$ and $\tilde{D}$ using the relation:
\begin{equation}
\Sigma(x,y;t) = {1 \over 4} \int dz G^{-1} (x,z;t) \{D(z,y;t)  
+\tilde{D}(y,z;t)\}\equiv
{1 \over 4} \int d^dz \{ \tilde{D}(x,z;t)+ D(z,x;t) \}G^{-1} (z,y;t).
\end{equation}
Using the equations of motion for $G$ $D$ $\tilde{D}$ and $K$ one can then show
that $ {dN \over dt} = 0$.  Thus if we choose our initial condition on K to  
satisfy the constraint:
\begin{eqnarray}
&&K(x,y;t=0)={1 \over 4} G^{-1} (x,y;t=0) \nonumber \\
&&+{1 \over 4} \int d^dz_1 d^dz_2  \{\tilde{D} (x,z_1;0)+D(z_1,x;0) \} G^{-1}  
(z_1,z_2;0)\{ D(z_2,y;0)+\tilde{D}(y,z_2;0)\} \nonumber \\
\end{eqnarray}
then the conservation law (which is equivalent to eq. (\ref{kdot}) guarantees
that this constraint is observed at all times.

There is a definite advantage in these update equations over the TDHF ones,  
as we will see in
specific examples below. The equal time Green's function equations  lead to a  
spatially
local update (apart from gradients), which is quite an advance over the
previous (TDHF)
case which required non local updates as well as matrix inversions and
therefore is computationally more intensive. The small price one pays is to
increase the number of variables by one.

\subsection{ Covariant approach to mean field theory }

The covariant approach to mean field theory, arises from a self consistent
truncation of the Schwinger Dyson hierarchy of equations for the N-point
Green's
functions to just the first two equations: namely an equation for the one
and two
particle connected Green's Functions. This can be accomplished directly in
terms
of the Dyson equations (as we do here), or by introducing a quadratic
constraint
field in a  path integral approach and performing a loop expansion with
respect
to the constraint field \cite{fred}.

Starting from the Lagrangian
\begin{equation}
L= \int d^d x dt \; \{ {1\over 2} \partial_{\mu}\varphi \partial^{\mu}\varphi -
V[\varphi] \} \end{equation}
we want to  obtain the equations of motion for the
expectation values  \begin{equation}
  \phi_c(\vec x,t) = Tr \rho \varphi(\vec x,t) = < \varphi(\vec x,t) >.
\end{equation}
and the  two time Wightman functions : $G^{<,>}(\vec x,\vec y;t,t')$ given by 
\begin{eqnarray}
 G^>(\vec x,\vec y;t,t')  & = &  Tr [\rho  \varphi(\vec x,t) \varphi(\vec
y,t')] - \phi_c(\vec x,t)
\phi_c(\vec y,t') \nonumber \\
G^<(\vec x,\vec y;t,t')  & = &  Tr [\rho  \varphi(\vec y,t)
\varphi(\vec x,t')] -
\phi_c(\vec y,t) \phi_c(\vec x,t')
\end{eqnarray}
 Expectation values are in the {\em initial} density matrix $\rho$.  The
various Green's Functions can be constructed from the two Wightman functions.
 In  the mean field
approximation, the exact equation of motion for $\phi_c$ is:

\begin{equation}
  \partial^2  \phi_c(x,t)+ < {\partial V[\varphi] \over \partial
\varphi (x,t)} >
  =0
\end{equation}
which is  approximated by replacing  the exact expectation value by the  
expectation value
using the Gaussian trial wave-functional.
Making the same approximation to the exact equation for the two point function
one obtains for the  Green's function:

\begin{equation}   \{ \partial^2 + < {\partial^2 V[\varphi] \over (\partial
 \varphi (x,t))^2 } >  \} G(\vec x,\vec y;t, t')  = \delta^d(\vec x-\vec y)
\delta_{\cal C}(t-t')
\end{equation}
Here because the boundary conditions pertain to an initial value problem,
the delta function is defined on the Closed Time Path (CTP) contour of
the Schwinger-Keldysh formulation\cite{ctp}.  Namely
\begin{equation}  \delta_{\cal C}(t-t') = {d \Theta_{\cal C} (t,t')
\over dt}
\end{equation}
where

\noindent The expectation value is obtained using the Gaussian
density matrix. Since the expectation values obtained in the Gaussian trial
state depend only on $ \phi_c$ and $G$ the system of coupled Green's function
equations truncates at this level. The causal Green's function is related to
the Wightman functions defined above as follows:
\begin{equation}
G(x,y,t,t') = \Theta_{\cal C}(t,t')  G^>(x,y,t,t')+\Theta_{\cal C}(t',t)
G^<(x,y,t,t').
\end{equation}
 In
the covariant approach one needs two time information about the Green's
function.
Thus if we directly solve this equation, starting from the initial
data at $t=t'=0$,  it would be
computationally much more storage intensive than considering the set of coupled
equal time Green's functions.  One way to solve  this Green's function
equation and simultaneously handle the mass renormalization problem is to
convert the partial differential equation into an integral equation. Defining  
$M^2(x,t)$ via
\[
M^2(x,t) \equiv < {\partial^2 V[\varphi] \over (\partial
 \varphi (x,t))^2 } >
\]
and calling $m_r^2$ the mass in the Hartree approximation in the vacuum sector,
then we have that
\begin{equation}
G(x,y,t,t') = G_0(x,y,t,t') - \int d^dz \int_{\cal C}dt'' G_0(x,z,t,t'')\Delta  
M^2(z,t'') G(z,y,t'',t')
\end{equation}
where
\[ \Delta M^2(x,t)) = M^2(x,t) -m_r^2 \]
and
\[ \int_{\cal C} dt = \int_{0, {\cal C}_+} ^{\infty}dt -
\int_{0, {\cal C}_-} ^{\infty}dt \]
The CTP formalism\cite{ctp} guarantees causality so that $ t''
<  \{ t, t' \}$.
Solving this integral equation numerically
is quite storage intensive. Here $G_0$ is
the free field theory  CTP Green's function for a particle with mass $m_r^2$.
Integral equations of this type and their numerical solution are discussed in
 \cite{bog}.

To circumvent these storage problems, one can use another strategy for
updating $G$ which relies on
finding a complete set of solutions for the quantum fluctuation field out of
which the Green's function can be constructed.
That is if we have a complete orthonormal set of solutions $\psi_k(x,t)$ to the
equation:

\begin{equation}   \{ \partial^2 + M^2(x,t)   \}\psi_k(x,t)  = 0,
\label{eq:mod}
\end{equation}
 where the scalar product is defined by
\begin{equation}
< f, g> =  i \int d^dx \; \left[ f^*(x,t) {\partial g(x,t) \over
\partial t} -g(x,t)
{\partial f^*(x,t) \over \partial t} \right] \; ,
\end{equation}
then we can expand the fluctuation field $ \hat{\psi} $ (where  $ \varphi
= \phi_c +  \hat{\psi} $) as follows:
\begin{equation}
\hat{\psi}(x,t) = \sum_k \{ a_k \;\psi_k(x,t) + a_k^{\dag} \;\psi_k^*(x,t)
 \} \end{equation}
with the canonical momentum $\pi(x,t)$ given by:
\begin{equation}
\pi (x,t) = \sum_k \{ a_k \;\partial_t \psi_k(x,t) + a_k^{\dag} \;\partial_t
\psi_k^*(x,t)
 \} \end{equation}
(for continuum modes $\sum_k \equiv \int {d^dk \over (2 \pi)^d} $).

The mode functions satisfy the orthogonality conditions:
\[ < \psi_k, \psi_l> = \delta(k-l)~~ ; ~~ < \psi_k^*, \psi_l> =0 \; . \]
Assuming the usual commutation relations for the creation and annihilation
operators:
\begin{equation}
[a_k, a^{\dag} _l] = \delta_{kl} ;~~ [a_k,a_l]= [a^{\dag}_k, a^{\dag}_l] =0\;.
\end{equation}
(where $\delta_{kl} \equiv (2 \pi)^d \delta^d(k-l)$ for continuum modes)
leads to the closure or completeness condition:
\begin{equation}
 i \sum_k  \left[ \psi_k^*(x,t) {\partial \psi_k (y,t) \over \partial t}
-\psi_k(x,t) {\partial \psi_k^*(y,t) \over \partial t} \right]= \delta(x-y)\;.
\label{eq:close} \end{equation}
\noindent One can verify using the equation of motion (\ref{eq:mod})  for the
modes, that  if the  modes are orthogonal at $t=0$ that they stay orthogonal at
all times.

For the case of spatial homogeneity, it is quite simple to find a
complete set of orthonormal solutions to the classical field problem at  all
times\cite{PRL96}-\cite{tsunami}. The mode equations
become for spatially homogeneous problems
\begin{equation}
\{ \partial^2 + M^2(t) \} \psi_k(x,t) =0
\end{equation}
Letting
\[ \psi_k(x,t) = f_k(t)\; e^{ikx} \]
we find the $f_k$ obey the ordinary differential equation:
\[ \ddot{f}_k + \{k^2+M^2(t)\} f_k =0 \]
In terms of the $f_k$ one then has that the fluctuation field can be
constructed at all times as:
\begin{equation}
\hat{\psi}(x,t) = \sum_k \{ a_k\; f_k(t)\; e^{ikx} + a_k^{\dag}\;
f_k^*(t)\; e^{-ikx} \}
\end{equation}
The closure relationship is then satisfied as long as $f_k$ obey the Wronskian
condition at $t=0$:
\[  f_k {d f_k^* \over dt}- f_k^* {d f_k \over dt}= i \]
Renormalization constraints require that for large values of $k$ the modes
$f_k$  differ in a prescribed fashion from the free field ones. The
renormalization can be performed by studying  the WKB
expansion of the mode functions and isolating the divergences as discussed
in several references: \cite{wkb,frw1}. The initial values of $f_k$ and
$\dot{f_k}$
are usually chosen so that the initial state is either in an adiabatic vacuum
state or a thermal state with respect to the adiabatic vacuum as discussed
in  \cite{wkb,frw1}.

  All the
non-equilibrium  Green's functions, can be constructed in terms of the Wightman
function
 \begin{equation}
G^> (\vec x,\vec y,t, t') = < \hat{\psi}(\vec x,t) \hat{\psi}(\vec y,t') >
 \end{equation}
in the usual fashion.

In the spatially inhomogeneous case, the mode functions obey the partial
differential equation:
\begin{equation}
\{ \partial^2 + M^2(x, t) \} \psi_k(x,t) =0
\end{equation}
 $M^2(x,t)$ is determined from the same modes that solve this equation, thus
it is necessary, in order to get started at $t=0$
to find a self consistent choice for $M^2(x,t=0)$
and the mode functions. This will allow us to determine the initial conditions
for the partial differential equations for each mode. This aspect of the
problem  will be addressed in the following sections.
 The second problem in this approach
is to find efficient symplectic integration methods which will allow
us to simultaneously update many partial differential equations, each of
which is similar to updating the time dependent Schr\"odinger equation.
The way that the initial conditions are handled is by assuming at $t=0$
that we can write:
\[
\psi_k(x,t) = {e^{-i\omega_k t} \over (2 \omega_k)^{1/2} }\; g_k(x);
~g_k^*=g_{-k} \] Then the closure relationship becomes the completeness
relation: \begin{equation}
\sum_k g_k^*(x) g_k(y) = \delta(x-y)
\end{equation}
and the $g_k$ obey a Schr\"odinger equation.
\begin{equation}
\{-\nabla^2 + M^2(x,t=0) \} g_k(x) =  \omega_k^2 \;g_k(x).
\end{equation}

The covariant approach to mean field theory is often the first term in a
systematic expansion of the full field theory, so that the advantage here
is that
one can determine the next order corrections and see how long the mean field  
approximation is valid.  This has been carried out in a simple quantum
mechanical
example in \cite{bog}.  The standard method for doing the systematic
expansion is to start with the path integral and use closed time path boundary
conditions to insure causality. This is discussed in great detail in
\cite{losalamos,largeN}. The advantage of having a mode function
decomposition for the Green's functions is that subtractions necessary
for renormalization can be made on a mode by mode basis, which dramatically
improves the convergence of the mode sums.

\subsection{Operator and Closed Time Path Methods: }

A convenient implementation of the Gaussian variational approximation
at the operator level begins with the Lagrangian formulation,
 shifting the field by its expectation value and implementing a
Hartree-like factorization of non-linear interaction
terms in order to make the Lagrangian {\em linear plus quadratic}. The
linear terms in the fluctuation are required to vanish, leading to
an equation of motion for the expectation value, the remaining Lagrangian
 is a quadratic form in terms of the fluctuations. Passing
to a Hamiltonian formulation for the operator fluctuations the Heisenberg
equations of motion for the operators is obtained, and the expectation
values are now understood in the Heisenberg representation in the initial
state.

 Specifically, for a Lagrangian density of the form

\begin{equation}
{\cal L}[\varphi] = \frac{1}{2}\partial_{\mu}\varphi\partial^{\mu}\varphi-
V(\varphi) \label{lagradens}
\end{equation}
(where we have omitted the linear coupling to sources), the Gaussian
approximation is implemented in the following steps:
i) shift the field $\varphi$ and its expectation value
\begin{eqnarray}
\hat{\varphi}(\vec x,t)  & = &    \varphi_c(\vec x,t) +
\hat{\psi}(\vec x, t)
\label{shift}\\
\langle \hat{\psi}(\vec x, t) \rangle & = &   0
\label{nullexp}
\end{eqnarray}

ii) After this shift the potential has in general even and odd powers of
the fluctuation $\psi$,
\begin{eqnarray}
&& V[\varphi_c + \psi] = V^{e}[\varphi_c,\psi]+ V^{o}[\varphi_c,\psi]
\label{evenodd} \\
&& V^{e}[\varphi_c,\psi] = V[\varphi_c]+ \psi^2 \left(\frac{\partial^2
V(\varphi)}{\partial\varphi^2}\right)_{\varphi=\varphi_c+\psi}
\label{evenpart} \\
&&V^{o}[\varphi_c, \psi] = \psi \left(\frac{\partial
V(\varphi)}{\partial \varphi}\right)_{\varphi=\varphi_c+\psi}
\label{oddpart}
\end{eqnarray}
The Hartree factorization replaces the even and odd parts by

\begin{eqnarray}
&&V^{e}_H[\varphi_c,\psi] \rightarrow V[\varphi_c]+ \psi^2 M^2(\vec
x,t)+ {\cal C}(\vec x,t) \nonumber \\
&& M^2(x,t) = \langle \frac{\partial^2
V(\varphi)}{\partial\varphi^2}|_{\varphi=\varphi_c+\psi} \rangle
\label{hartevenpart} \\
&&V^{o}_H[\varphi_c, \psi] = \psi \; {\cal V}(\vec x,t)\nonumber\\
&&{\cal V}(\vec x,t) = \langle
\frac{\partial V(\varphi)}{\partial \varphi}|_{\Phi=\varphi_c+\psi}
\rangle \label{hartoddpart}
\end{eqnarray}

Where the expectation values will be obtained self-consistently. The
field independent term   ${\cal C}$ is obtained by requesting that the
expectation value of the potential equals the expectation value of its
Hartree factorized form.

Upon integration by parts in the resulting action, the linear term in
the fluctuation $\psi$ is requested to vanish to ensure the condition
(\ref{nullexp}), leading to the equation of motion for the expectation
value

\begin{equation}
{\partial^2 \varphi_c \over \partial t^2}-\nabla^2\varphi_c + {\cal V}(\vec x,t) =0 
\label{classicaleqn}
\end{equation}

Passing to the Hamiltonian in terms of the operators $\hat{\psi}$  and
their canonical momenta $\hat{\pi}$

\begin{equation}
H = \frac{1}{2}\int d^dx \left[ \hat{\pi}^2 +
(\nabla\hat{\psi})^2+M^2(\vec x,t)\hat{\psi}^2+{\cal C}(\vec x,t)
\right] \label{hamiop}
\end{equation}

the Heisenberg equations of motion for the operators become

\begin{equation}
\left[{\partial^2  \over \partial t^2}- \nabla^2 + M^2(\vec x,t)
\right]\hat{\psi}(\vec x,t) =0 \label{heiseneqn}
\end{equation}

The Hartree factorization requires the quantity
\begin{equation}
G(x,x;t) = \langle \hat{\psi}^2(\vec x,t) \rangle \label{gxx}
\end{equation}
in which $<\cdots>$ stands for expectation value in the initial
Gaussian wave-functional  or density matrix since in the Heisenberg
picture the operators evolve in time but the states do not. Since
${\cal V}$ and $M^2$ are expectation
values of the first and second derivative of the potential as given
by eqs. (\ref{hartevenpart},\ref{hartoddpart}) this implies a functional
relation of both ${\cal V} \; ; \; M^2$ on $G(x,x;t)$ leading to
a self-consistent condition.

An alternative description of the non-equilibrium dynamics is based
on the Keldysh or closed time path method\cite{ctp} in which the main  
ingredient is the time evolution of an initial density matrix, which could  
describe either a pure or mixed state
\begin{equation}
\hat{\rho(t)} = U(t,t_0) \hat{\rho(t_0)}U^{-1}(t,t_0)
\label{rhooft}
\end{equation}
where $U(t,t_0)$ is the time evolution operator, which is completely
determined by the Hamiltonian of the system, with boundary condition
$U(t_0,t_0)=1$.

Non-equilibrium expectation values and correlation functions in this
time evolved density matrix can be handily obtained via a path integral
representation of the generating functional of non-equilibrium correlation  
functions along a contour in complex time\cite{ctp}
\begin{equation}
Z[j^+,j^-] = \int {\cal D}\varphi^+ {\cal D} \varphi^- e^{i \int dt d^dx  
\left\{ {\cal L}[\varphi^+,j^+] - {\cal L}[\varphi^-,j^-]  \right\}}
\label{cpt}
\end{equation}
where the fields and sources $\varphi^{\pm} \; ; \; j^{\pm}$ refer to the
forward (+) and backward (-) time branches, and functional derivatives
with respect to the sources generate all the non-equilibrium correlation
functions. The reader is referred to the ample literature for more
details\cite{ctp,boysinglee,largeN,erice97}.

The Gaussian approximation is implemented by performing the shift of
fields given by eq. (\ref{shift})  with the same
$\varphi_c$ on both branches since it is the expectation value and with
the factorization given by
eqs. (\ref{hartevenpart},\ref{hartoddpart}) for the fields on both
branches.

 In this factorization scheme and after an integration by parts in the
 generating functional, the effective lagrangian density becomes

\begin{equation}
{\cal L}[\varphi_c, \psi] = {\cal L}[\varphi_c]+\frac{1}{2} \left\{
\partial_{\mu}\psi\partial^{\mu}\psi - \psi^2 M^2 \right\} + \psi
\left\{ -\partial_{\mu}\partial^{\mu}\varphi_c - {\cal V}\right\}
\label{quadlagra}
\end{equation}
The effective Lagrangian density necessary for the calculation of
non-equilibrium correlation functions in the CTP-Keldysh scheme is
given by ${\cal L}_{eff}[\varphi^+,\varphi^-] = {\cal L}[\varphi_c,
\psi^+]-{\cal L}[\varphi_c, \psi^-]$.
Thus we see that this factorization scheme leads to a Gaussian
generating functional in which the correlation functions are obtained
self-consistently. Examples are described in detail below.
The condition  (\ref{nullexp}) is ensured by requiring the vanishing
of the term linear in $\psi$. This is the tadpole
equation\cite{erice97} that leads to the equation of motion for the
expectation value given by (\ref{classicaleqn}).

An alternative manner to confirm the Hartree factorization given by
eqs. (\ref{evenodd},\ref{evenpart},\ref{oddpart}) in this scheme is to
write in the original Lagrangian density the even and odd parts of the
potential after the shift (\ref{shift}) as

\begin{eqnarray}
&&V^{e}[\varphi_c,\psi] = V[\varphi_c]+ \psi^2 M^2(\vec x,t)
+\delta^e \label{deltae} \\
&&V^{o}[\varphi_c, \psi] = \psi {\cal V}(\vec x,t)+ \delta^o \label{deltao}
\end{eqnarray}
and treat $\delta^{e,o}$ as `counterterms'. The Gaussian approximation
is completely determined by the one and two point functions, and
these counterterms are obtained by requiring that $\delta^o$ gives
a vanishing contribution to the one point function and $\delta^e$ gives
a vanishing contribution to the two point function. This requirement
leads to the identifications of ${\cal V}$ and $M^2$ given in eqs.
(\ref{hartevenpart},\ref{hartoddpart}).

In the CTP formulation  the important non-equilibrium Green's
functions are given by the time-ordered ${\cal G}^{++}$, anti-time
ordered ${\cal G}^{--}$ and Wightman ${\cal G}^{+-}\; , \; {\cal G}^{-+}$
functions given by

\begin{eqnarray}
{\cal G}^{++}(x,t;x',t') & = &  \langle \psi^+(x,t)\psi^+(x',t')
 \rangle =
\langle T \psi(x,t)\psi(x',t')\rangle
\label{g++} \\
{\cal G}^{--}(x,t;x',t') & = &  \langle \psi^-(x,t)\psi^-(x',t')
 \rangle = \langle \tilde{T} \psi(x,t)\psi(x',t')\rangle
\label{g--} \\
{\cal G}^{+-}(x,t;x',t') & = &  -\langle \psi^+(x,t)\psi^-(x',t')
 \rangle = -\langle \psi(x',t')\psi(x,t) \rangle
\label{g+-} \\
{\cal G}^{-+}(x,t;x',t') & = &  -\langle \psi^-(x,t)\psi^+(x',t')
 \rangle = -\langle \psi(x,t)\psi(x',t') \rangle
\label{g-+}
\end{eqnarray}
With $T \; ; \; \tilde{T}$ the time and anti-time ordering operators
respectively.

  Since the Hartree approximation is local in space and time
 ${\cal V}$ and $M^2$ are local in space and time, the self-consistent
dynamics only involves  the spatial coincidence limit of the
following Wightman function
\begin{equation}
G(\vec x,\vec x';t)=\langle \hat{\psi}(\vec x, t)
\hat{\psi}(\vec x', t)\rangle = -{\cal G}^{+-}(x,t;x',t)=-
{\cal G}^{-+}(x,t;x',t)\label{G}
\end{equation}

The time evolution of this Green's function leads to a closed Hamiltonian
system of equations equivalent to the method of equal time Green's functions
presented above, eqns. (\ref{ficeq}-\ref{kdot}) with the identification of the self consistent 
mass $M$ and ``potential''  ${\cal V}$:

\begin{eqnarray}
 && \dot{G}(\vec x, \vec x';t)   =
D(\vec x, \vec x',t)+\tilde{D}(\vec x, \vec x',t)
 \label{firsteqn} \\
&& D(\vec x, \vec x',t)  =   \langle \hat{\dot{\psi}}(\vec x, t)
\hat{\psi}(\vec x', t)\rangle \; ; \;
\tilde{D}(\vec x, \vec x',t) =\langle \hat{{\psi}}(\vec x, t)
\hat{\dot\psi}(\vec x', t)\rangle = D(\vec x', \vec x,t)+
i\delta^d(\vec x-\vec x') \label{Gdot} \\
 && \dot{D}(\vec x, \vec x',t)  =   K(\vec x, \vec x',t)+
\left(\nabla^2_x -M^2(\vec x,t)\right)G(\vec x,\vec x',t) \nonumber \\
 && \dot{\tilde{D}}(\vec x, \vec x',t)  =   K(\vec x, \vec x',t)+
\left(\nabla^2_{x'} -M^2(\vec x',t)\right)G(\vec x,\vec x',t)
\label{thirdeqn} \\
&& K(\vec x, \vec x',t)  =   \langle \hat{\dot{\psi}}(\vec x, t)
\hat{\dot{\psi}}(\vec x', t) \rangle \label{Ddot} \\
&& \dot{K}(\vec x, \vec x',t) =  \left(\nabla^2_x-M^2(\vec x,t)\right)
D(\vec x,\vec x',t)+\left(\nabla^2_{x'}-M^2(\vec x',t)\right)
\tilde{D}(\vec x,\vec x',t) \label{kkdot}
\end{eqnarray}

 This system of equations is Hamiltonian and can be updated in time by
providing initial Cauchy data on the three functions $G,D,K$ at an
initial time as a function of {\em two} variables $\vec x, \vec x'$.
Since ultimately we only need the concidence limit to solve the
self-consistent dynamical problem not all of the range of $\vec x'$ is
necessary
to be stored numerically and for a given value of $\vec x$ only values
of $\vec x'$ in the immediate neighborhood are necessary (see
discussion in section VII).


Of all the different available formulations that we presented above,
we prefer the closed-time path method with the Hartree factorization
leading to the evolution equations (\ref{firsteqn}-\ref{kkdot}) as these
present an economical and implementable system of update equations. However the main purpose of enumerating and analyzing the different alternatives is to provide  a detailed assessment of the advantages
and shortcomings of each method. 

However before deciding on a numerical implementation of this system of
equations two important steps are needed, the first is to provide
self-consistent initial data for $G,D,K$ and second is to address the
renormalization issues to guarantee that the dynamics will not depend
on ultraviolet (short distance) cutoffs. We will address these two issues
in definite and relevant $1+1$ and $3+1$ dimensional examples below.

\section{Initial Value Problem}

In all of the methods detailed in the previous section in which a
closed Hamiltonian system
of equations for the Green's functions determines completely the
dynamics, the first step
towards a dynamical implementation is to provide the initial value
problem, i.e. the Cauchy
data at the initial time for the Green's functions.


The time evolution will be completely determined upon providing initial
conditions on $\varphi_c$ and the Green's functions. The self-consistent
approach requires to solve self-consistently the initial condition problem at $t=0$. Once the solution to the self-consistent initial value
problem is obtained the self-consistent time evolution is obtained from
the closed time path set of equations (\ref{firsteqn}-\ref{kkdot}) obtained in the previous section.

 Because the set of
equations are
self-consistent, to provide this initial data is tantamount to solve
the initial value
problem for the Green's functions self-consistently for an initial
inhomogeneous classical
field configuration $ \varphi_c(\vec x,t=0) $, where  we have chosen
$t=0$ as the initial time.

Our  strategy for setting up the initial value problem and the Cauchy
data for the Green's function is more transparently described by
focusing on the Heisenberg equations (\ref{heiseneqn}).
At the initial time we expand the quantum fluctuation field $\psi$
 and its canonical conjugate momentum $\hat{\pi}= \hat{\dot{\psi}}$ in
terms of a complete set of functions as

\begin{eqnarray}
&& \hat{\psi}(\vec x,t=0)  =  \int  \frac{d^d k}{\sqrt{2\omega_k}}\left\{
a_{\vec k}\; \psi_k(\vec x)+ a^{\dagger}_{\vec k}\; \psi^*_k(\vec x) \right\}
\label{psiexpansion} \\
&&\hat{\pi}(\vec x,t=0)= \hat{\dot{\psi}}(\vec x,t=0)  =  -i \int
 d^d k\; \sqrt{\frac{\omega_k}{2}} \left\{ a_{\vec k} \;\psi_k(\vec x)-
a^{\dagger}_{\vec k}\; \psi^*_k(\vec x) \right\} \label{piexpansion}
\end{eqnarray}
with $ a_{\vec k} \; ; \; a^{\dagger}_{\vec k} $ canonical annihilation and
destruction operators. The
mode functions $\psi_k(\vec x)$ are chosen to be the complete set of
solutions of the following eigenvalue problem at the initial time

\begin{equation}
\left[-\omega^2_k -\nabla^2  + M^2(\vec x,t=0) \right]\psi_k(\vec x) =
0 \label{modeeqn}
\end{equation}

\noindent In terms of these modes and the expansion of the quantum
field and its canonical momentum given by
eq. (\ref{psiexpansion},\ref{piexpansion}) the Hamiltonian
(\ref{hamiop}) at the initial time ($t=0$) is given by
\begin{equation}
H(t=0) = \int d^dk \; \omega_k (a^{\dagger}_k a_k +1/2)+ \int d^dx \; {\cal
C}(\vec x,t=0) \label{hamnum}
\end{equation}

\noindent Having found the complete set of mode functions, the initial
values of $G(\vec x, \vec x';t=0)
\; ; \; D(\vec x, \vec x';t=0) \; ; \; K(\vec x, \vec x',t=0)$ are
completely determined
by determining the initial correlations $\langle a^{\dagger}_k
a_{k'}\rangle \; ; \;
\langle a_k a_{k'} \rangle$ in the initial state. The most  physically
reasonable
choice is given by
\begin{equation}
\langle a^{\dagger}_k \; a_{k'}\rangle = N_k \; \delta_{k,k'} \; ; \;
 \langle a_k \; a_{k'} \rangle =0 \label{inidistri}
\end{equation}

This choice corresponds to an initial state that is a density matrix
which is diagonal in the basis of the number operator $a^{\dagger}_k a_k$
with occupation number $N_k$. In particular for $N_k=0$ the initial
state is a pure state and corresponds to the Fock vacuum annihilated
by the $a_k$ and is given by the instantaneous ground state of the
Hamiltonian (\ref{hamiop}) at $t=0$.

At this stage we can make contact with the variational wave-functional
by expanding the
operator $\psi$ in the Schr\"odinger representation in terms of the
complete set of states
$\psi_k(\vec x)$ at $t=0$
\begin{equation}
\psi(\vec x,t=0) = \int d^d k \; \eta_k \; \psi_k(\vec x) \quad ; \quad \eta^*_k =
\eta_{-k} \label{expa}
\end{equation}
where he choose the mode functions in such a way that $\psi^*_k(\vec
x) = \psi_{-k}(\vec x)$
and the condition on the coefficients $\eta_k$ in (\ref{expa}) is a
result of the hermiticity of the field $\psi(\vec x)$. A pure state
wavefunctional at the initial time
can now be written  as
\begin{eqnarray}
\Psi[\eta] & = & N \; e^{ - \sum_{k,k'}K_{k,k'} \eta_k
\eta_{k'}+i\sum_k \pi_{c,k} \eta_k } \nonumber  \\
K_{k,k'} & = & \int d^d\vec{x} \; d^d\vec{x'} \; {\cal K}(\vec x,\vec
x') \;\psi_k(\vec x)  \; \psi_{k'}(\vec x') \nonumber \\
\pi_{c,k} & = & \int d^d \vec x  \;\pi_c(\vec x) \; \psi_k(\vec x)
\label{fourimom}
\end{eqnarray}

The choice leading to the initial correlations (\ref{inidistri}) corresponds to
a wave-functional which is the ground state for all of the harmonic modes with
frequencies $\omega_k$ and $\pi_c=0$, i.e.
\begin{equation}
K_{k,k'} = \frac{\omega_k}{2}\;  \delta_{k,-k'} \; \; ; \; \; \pi_{c,k} =0
\label{inikerkpi}
\end{equation}

This choice leads to the identification of the real  and imaginary parts of  
the kernel
\begin{equation}
G^{-1}_{k,k'} = 2\omega_k \delta_{k,k'} \;\; ; \; \; \Sigma_{k,k'}=0  
\label{iniGsigma}
\end{equation}

For an analogous approach see\cite{baackerothe}.

\section{$1+1$ Dimensions: Kinks and Lumps}

We begin by implementing the strategy to set up the initial Cauchy
data as described in the previous section in two relevant examples in
$1+1$ dimensions. Not only  does $1+1$ dimensions provide a simpler setting
within
which to gain experience and intuition, but also it is a relevant example
for statistical mechanics and condensed matter physics since both Sine-Gordon
and $\phi^4$ theories are models for quasi-one dimensional charge
density wave systems\cite{schrieffer,yu,gruner}. In these systems
transport phenomena is mediated by
coherent and collective excitations such as solitons, breathers and
pulse-like excitations known as polarons\cite{schrieffer,yu,gruner}.
Furthermore there is a large body of results on these quantum field
theories\cite{rubinstein}-\cite{sGCuan} that we
can use as a yardstick against which to compare our approximations

\subsection{Sine-Gordon field theory}

The exactly integrable Sine-Gordon field theory provides a good example to  
test the reliability of variational methods
since  the quantum S-matrix is exactly known for all particles as well as the
expectation values of local operators on asymptotic states.
This theory  is also known to have a phase transition as the coupling
constant is  increased\cite{coleman,amit}.

The Sine-Gordon field theory is described by the Hamiltonian:
\begin{equation}
H= \int dx [ {1 \over 2} \pi(x,t)^2 + {1\over 2} (\nabla \varphi(x,t))^2
- {\alpha_0 \over \beta^2} \cos \beta \varphi  ]
\end{equation}

Following the steps described in detail in the previous sections, we find  
that the Hartree factorization for this problem leads to
\begin{eqnarray}
{\cal V}(x,t)& = & \frac{\alpha_0}{\beta}  
e^{-\frac{\beta^2}{2}G(x,x;t)}\sin\beta\varphi_c(x,t)  \label{poten} \\
M^2(x,t) & = & {\alpha_0}e^{-\frac{\beta^2}{2}G(x,x;t)}
\cos\beta\varphi_c(x,t) \label{singormass}\\
G(x,x;t) & = & \langle \hat{\psi}^2(x,t) \rangle \label{selfsingor}
\end{eqnarray}

The equations of motion that determine the dynamics in the Gaussian
approximation thus become
\begin{eqnarray}
&& {\partial^2 \varphi_c  \over \partial t^2} -{\partial^2 \varphi_c  \over  
\partial x^2}+  
\frac{\alpha_0}{\beta}e^{-\frac{\beta^2}{2}G(x,x;t)}\sin\beta\varphi_c =0
\cr \cr
&& \left[{\partial^2  \over \partial t^2} -{\partial^2  \over \partial x^2}+  
{\alpha_0}
e^{-\frac{\beta^2}{2}G(x,x;t)}\cos\beta\varphi_c \right]\hat{\psi}(x,t)
 = 0 \label{flucsingord}
\end{eqnarray}
with the self-consistency condition given by eq.(\ref{selfsingor}).

\subsubsection{\bf `Vacuum' sector:}

As mentioned above, the Sine-Gordon field theory has a phase
transition at a critical value of the coupling constant\cite{coleman,amit}.
We will now show that this Gaussian approximation reproduces this results of  
the Sine-Gordon field theory, giving us confidence in the
approximation. For this we consider the vacuum sector with a homogeneous  
value of the expectation value $\varphi_c \equiv 0 $ modulo $2\pi$.
In this simple case the Heisenberg equations are solved in terms of plane  
waves, leading to
\begin{eqnarray}
G_0(x,y) & = & \frac{1}{2\pi} \int dk g_0(k) e^{i k(x-y)} \nonumber \\
 g_0(k) & = & \frac{1}{2(k^2 + \mu^2)^{1/2}}\; ;
\; ~ \mu^2 =  \alpha_0 ~ e^{- {\beta^2 \over 2} G_0} \; ; \; G_0=G_0(x,x)  
\label{freeGF}
\end{eqnarray}
 In the vacuum sector we find for the expectation value of the Hamiltonian  
density in the Gaussian state

\begin{equation}
{\cal{H}}(\mu^2) = {1 \over 8 \pi} \int_{-\Lambda}^{\Lambda} dk
{2 k^2 + \mu^2 \over (k^2 + \mu^2)^{1/2} } - {\alpha_0 \over \beta^2}
 ( {\Lambda^2
\over \mu^2})^ {-\beta^2/ 8 \pi}
\end{equation}
where we have introduced an upper momentum cutoff in the $k$-integral.
We now define the renormalized mass parameter
\begin{equation}
\alpha_r = \alpha_0 ({m^2 \over \Lambda^2})^{\beta^2/ 8 \pi}
\end{equation}
with $G_0=\langle \hat{\psi}^2 \rangle$ in the vacuum state and $\kappa^2$ an  
arbitrary renormalization scale.
The  energy density once subtracted at the renormalization scale $\kappa^2$  
is given by
\begin{equation}
{\cal{H}}(\mu^2)-{\cal{H}}(\kappa^2)= {1 \over 8 \pi} (\mu^2 - \kappa^2)
- {\alpha_r \over \beta^2} ({\mu^2 \over \kappa^2})^{\beta^2 / 8 \pi}
\end{equation}
Choosing the subtraction point to be
\[   \kappa^2 = \alpha_r \]
then the extremum  of the energy is at
\[  \mu^2 = \kappa^2. \]
Since the second derivative of the energy is given by:
\begin{equation}
{\partial^2 {\cal{H}} \over (\partial \mu^2)^2} = -{1 \over 8 \pi \kappa^2}
({\beta^2 \over 8 \pi}-1) ({\mu^2 \over \kappa^2})^{(\beta^2/ 8 \pi)-2}
\end{equation}
we find that a stable ground state exists only for
\begin{equation}
{\beta^2 \over 8 \pi} < 1
\end{equation}
which is in accord with the result of Coleman\cite{coleman}. For
$\beta^2 > 8\pi$ the Sine-Gordon theory undergoes a Kosterlitz-Thouless
phase transition\cite{amit}. Whereas the theory is superrenormalizable
for $\beta^2 <8\pi$ it becomes renormalizable at $\beta^2 > 8\pi$
since the
cosine operator becomes marginal at this value of
$\beta$\cite{amit}. The region $\beta^2 >8\pi$ must be studied within
a double expansion in $\alpha_r$ and $\delta=\beta^2-8\pi$\cite{amit}.

 We will restrict our study of the Sine-Gordon field theory to the
 superrenormalizable regime i.e. $\beta^2 < 8\pi$ since the Gaussian
 approximation cannot describe correctly the physics for $\beta^2 >
 8\pi$\cite{amit}.

\subsubsection{\bf Initial value problem: Self-consistent Kinks and Lumps}

At the classical level,  the Sine-Gordon
model has moving kink solutions\cite{rubinstein}
\begin{equation}
\phi(x,t) = {4 \over \beta} \tan^{-1}\left\{ \exp \left[ {\alpha_0
\over (1-v^2)^{1/2}}
(x-q_0-vt) \right]\right\} \label{eq:clkink}
\end{equation}
where $q_0$ is the position of the kink at time $t=0$ and
$ v $ its velocity. The dependence  of the solution on $ x, t $
reflects the translational and Lorentz invariance of the theory.
The energy of this moving kink is\cite{rubinstein,rajaraman}
\begin{equation}
E = {8 \; \alpha_0 \over (1-v^2)^{1/2}}
\end{equation}
Thus the classical (c-number) kink is a relativistically invariant
model of an extended body in one spatial dimension.

For a kink solution at rest we have:
\begin{eqnarray}
\cos \beta \phi(x) && =  1-{2 \over \cosh^2 \alpha_0 x} \nonumber \\
\sin \beta \phi(x) && =  {2\sinh \alpha_0 x \over \cosh^2 \alpha_0 x}
\end{eqnarray}

We seek to set up an initial value problem that can include as  
self-consistent solutions kink-like configurations to be identified
as a classical kink dressed by the quantum corrections. As described
in a previous section, this is achieved by seeking self-consistent
solutions of the following eigenvalue problem

\begin{eqnarray}
&& \left[-\omega^2_k - \frac{d^2}{dx^2}+ \mu^2
e^{-\frac{\beta^2}{2}\tilde{G}(x,x)}\cos\beta\varphi_c(x)\right]\psi_k(x)=0
\label{eigensingor}\\
&& \tilde{G}(x,x) = G(x,x)-G_0
\end{eqnarray}
where $\mu^2\; ; \; G_0$ are given by eq.(\ref{freeGF}) and with
$\varphi_c$ {\em a} classical configuration with a kink-like
profile to be found self-consistently. In the
presence of a kink-like configuration the quantum fluctuations
asymptotically far away from the kink will be similar to vacuum
fluctuations since $\varphi_c(x) \rightarrow 2\pi$ far away from the
kink, therefore $G(x,x)-G_0 \rightarrow 0$ as $x \rightarrow \infty$.
This leads to the effective mass of the excitations to be given by $\mu$.

The spectrum of harmonic quantum fluctuations around soliton solutions
of the Sine-Gordon theory is
obtained from a Schr\"odinger like equation with a a potential
of the form
\begin{equation}
v(x) = m^2 \frac{n(n+1)}{\cosh^2[mx]} \label{integrable}
\end{equation}
where $m$ is the (asymptotic) mass of the fluctuations and for the
Sine-Gordon model $n=1$. Potentials of this form are reflectionless and
integrable  and
their spectrum is exactly known\cite{morse}. Therefore
we propose a self-consistent potential for the mode functions of the
form
\begin{equation}
\mu^2 e^{-\frac{\beta^2}{2}\tilde{G}(x,x)}\cos\beta\varphi_c(x) \equiv
M^2- \frac{2m^2}{\cosh^2[mx]} \label{ansatzsingor}
\end{equation}
where the parameters $M^2 \; ; \; m^2$ are identified with the asymptotic
mass of the quanta and the width of the kink or lump configuration
and will be self-consistently determined from the parameters of the
theory $\mu^2 \; ; \; \beta$.

We want to stress that eq.(\ref{ansatzsingor}) is an {\em additional}
ansatz, {\em independent} of the variational approximation defined by
eqs.(\ref{selfsingor})-(\ref{flucsingord}). We
remark that although
Eqs. (\ref{selfsingor})-(\ref{flucsingord}) admit much more general
solutions, the simple ansatz (\ref{ansatzsingor}) is introduced to facilitate  
an analytic self-consistent solution to the initial value problem.
Eqs.(\ref{selfsingor})-(\ref{flucsingord})  in general require a
numerical solution (see sec. VII) which is computationally intensive.

With this ansatz the eigenvalue problem becomes
\begin{equation}
\left[ \frac{d^2}{dx^2}+W^2_k -m^2
+\frac{2m^2}{\cosh^2[mx]}\right]\psi_k(x)=0 \; \; ; \; \; W^2_k =
w^2_k+m^2-M^2
\label{eigenprobSG}
\end{equation}
In terms of the solutions of this eigenvalue problem we will choose the
initial Gaussian state to be that of the ground state for the Harmonic
oscillators of frequency $\omega_k$.

The solutions to this eigenvalue problem are available in the  
literature\cite{rubinstein,morse}
\begin{eqnarray}
\psi_0(x) = \sqrt{\frac{m}{2}} \frac{1}{\cosh[mx]} \; ; \; W^2_k= 0  
\Rightarrow \; \omega^2_0 = M^2-m^2 \label{zeromodeSG} \\
\psi_k(x) = \frac{e^{ikx}}{\sqrt{k^2+m^2}}\left[ik-m\tanh[mx]\right]
\; ; \; W^2_k = k^2+m^2 \Rightarrow \omega^2_k = k^2+M^2 \label{kmodsSG}
\end{eqnarray}
We see that the mass of the asymptotic particle states around the classical  
configuration is given by $M$ allowing us to identify at
once this parameter with the (renormalized) mass parameter in the
Sine Gordon equation
\begin{equation}
M^2 = \mu^2 . \label{massSG}
\end{equation}

Requiring the frequency of the bound state (\ref{zeromodeSG}) to be
positive restricts $M^2 > m^2$, the continuum modes asymptotically
are phase shifted plane waves. The Heisenberg field operators and its
canonical conjugate momentum  at the initial time are expanded as
\begin{eqnarray}
\hat{\psi}(x,t=0) & = & \frac{1}{\sqrt{2\omega_0}}(a_0+a^{\dagger}_0)\;
\psi_0(x)+ \sum_k \frac{1}{\sqrt{2\omega_k}}\left[a_k \; \psi_k(x)+
a^{\dagger}_k \; \psi^*_k(x) \right] \label{opeSG} \\
 \hat{\pi}(x,t=0) & = &
-i\sqrt{\frac{\omega_0}{2}}(a_0-a^{\dagger}_0)\;
\psi_0(x)-i \sum_k \sqrt{\frac{\omega_k}2}\left[a_k\;
\psi_k(x)- a^{\dagger}_k \;\psi^*_k(x) \right] \label{piopeSG}
\end{eqnarray}

As mentioned above the initial state is chosen to be the ground state
for the Harmonic oscillators of frequency $\omega_k$, i.e. this state
is the vacuum of all the annihilation operators. We find
\begin{eqnarray}
G(x,x) & = & G_0 + \frac{F[\eta]}{\cosh^2[mx]} \; ; \; \eta =
\frac{m}{\sqrt{M^2-m^2}} \label{GofxSG} \\
F[\eta] & = & \frac{\eta}{2\pi}\mbox{ArcTg}[\eta] \label{F}
\end{eqnarray}

The ansatz (\ref{ansatzsingor}) leads to the following consistency
equations
\begin{eqnarray}
&& M^2 = \mu^2 \nonumber \\
&& \cos\beta\varphi_c(x) = e^{\frac{\beta^2 F[\eta]}{2\cosh^2[mx]}}
\left[1-\frac{2\eta^2}{1+\eta^2}\frac{1}{\cosh^2[mx]}\right]\label{ficons}
\end{eqnarray}
The first equation is simply the statement that the asymptotic states
far away from the soliton must describe quanta of mass $\mu$, which
has already been argued above in eq.(\ref{massSG}). For a topological
kink solution we must impose that $\varphi_c(x\rightarrow -\infty)  
\rightarrow 0 \; ; \; \varphi_c(x\rightarrow +\infty) \rightarrow 2\pi$.
This requires that $\cos\beta\varphi_c(x\rightarrow \pm \infty) \rightarrow  
1$ but it must cross zero twice when the argument is $\pi/2\; ; \; 3\pi/2$ and  
that the minimum value of the cosine is $-1$.
These two conditions lead to
\begin{eqnarray}
&& \eta^2 >1 \label{cond1} \\
&& \frac{\beta^2}{2} = \frac{1}{F[\eta]}\ln[\frac{\eta^2+1}{\eta^2-1}]
\label{cond2}
\end{eqnarray}
The second condition (\ref{cond2}) determines $m=m(\beta)$ and for all
values of $\beta$ it is  constrained  in the interval
\begin{equation}
\frac{\mu^2}{2} \leq m^2 \leq \mu^2 \label{constraiSG}
\end{equation}
this is a rather mild constraint on the width of the soliton and confirms
that in the weak coupling limit $\beta \rightarrow 0$ $m \rightarrow \mu  
\approx \alpha_0$ as is expected.  Within the region (\ref{constraiSG})
there are solutions for arbitrary $\beta$.

This method also allows us to construct self-consistent solutions of
the initial value problem for the case of {\em non-topological} lumps.
For all values of the parameters (still with $\mu^2 > m^2$) such that
\begin{equation}
e^{\frac{\beta^2}{2}F[\eta]}\left[\frac{\eta^2-1}{\eta^2+1}\right]<1
\label{lumpSG}
\end{equation}
the solution provides a classical field configuration whose asymptotic
value resides in the same vacuum state and is clearly an excited state
that will relax into mesons under time evolution.

In this variational approximation, physical quantities, such as the soliton
profile, the energy, etc, are analytic functions of $\eta^2$.
It is then interesting
to connect $\eta^2$ with the SG coupling constant $\beta$. In the weak
coupling limit ($ \beta \to 0 $), we see that  $ \eta^2 \to \infty
$. More precisely,  eqs.(\ref{F}) and (\ref{cond2}) yield,
$$
\eta \buildrel {\beta\to 0} \over = \left({4 \over \beta}\right)^{2/3}
\left[ 1 + {2 \over {3 \pi}}\left({\beta \over 4}\right)^{2/3}+
{\cal O}(\beta^{4/3})\right]
$$
Therefore, the variational approximation yields for weak coupling a
power series in $ \eta^{-2} \sim \beta^{4/3} $. This should be compared
with the exact results which can be expressed in power series of $
\beta^2 $ (besides essential singularities at $ \beta = 0 $). In other
words, the variational approximation {\em does not} reproduce the analytic
properties in the coupling constant since a cubic-root cut in  $ \beta
$ is suggested and the weak coupling theory and the variational approximation  
cannot be simply matched in a series in $\beta$.

\subsection{Comparison with exact results in the quantum sine-Gordon
theory}

The  expectation values of some interesting operators are
exactly known in one-particle (and multi-particle) states \cite{wk,smi}.
Among them  is $ \cos\beta\phi $ which is just the trace of the
energy-momentum tensor,
$$
 T_{\mu}^{\mu}(x) = {2 \alpha_0 \over {\beta^2}} [ 1 -
 \cos\beta\phi(x)] \; .
$$

Let us consider (quantum) one-soliton states $ | \theta > $
with rapidity $ \theta $. That is a physical state with energy and
momentum:
$$
E = M \cosh \theta \quad , \quad p = M \sinh \theta
$$
where $ M $ is the physical quantum soliton mass.

The expectation value of $ T_{\mu}^{\mu}(0) $ in such
states can be written as,

\begin{equation}\label{ff}
F(\theta) \equiv <\theta_1| T_{\mu}^{\mu}(0) |\theta_2>\; .
\end{equation}
Here $ \theta= \theta_1-\theta_2 $.

One has the exact formula\cite{smi},
\begin{equation}\label{ffexa}
F(\theta)= - {M \over {4 \; \xi}} {{\sinh\theta}\over
{\sinh{{\pi\theta}\over {2\; \xi}}}}\; e^{I(\theta)} \; ,
\end{equation}
where
\begin{eqnarray}
\xi &=& {\beta^2 \over {8 \left( 1 -  {\beta^2 \over {8 \pi}} \right)}}
\; \cr \cr
\mbox{and}\cr \cr\label{minim}
I(\theta)&=&\int_0^{\infty}{dx \over x}\;
{{\sinh\left[\frac12(\pi-\xi)x \right] \;
\sin^2\left[\frac12 x\theta\right] }\over {\sinh\left[\pi x\right]
\sinh\left[\frac12x\xi\right] \cosh \left[ \frac12\pi x\right] }}
\end{eqnarray}
[Our normalization convention for the soliton states differs on a
factor $ M $ from ref.\cite{smi}].

The integral in eq.(\ref{minim}) yields the so-called minimal form
factor\cite{wk}. It can be written as a double-infinite product of Gamma
functions. In this way one explicitly sees that $ F(\theta) $ is a
meromorphic function of $ \theta $.  The closest pole of $ F(\theta) $
to the real  $ \theta $-axis  is at $ \theta =\pm 2 i \xi $.

For large momentum transfer, the expectation value (\ref{ff}) exhibits
 Regge behavior;
$$
F(\theta)\buildrel {\theta >> 1} \over =  -
{t^{1- {2\pi \over \beta^2}} \over{4 \; \xi\; M^{1- {4\pi \over \beta^2} }}}
\; ,
$$
where $ t \equiv 4 M^2 \cosh^2 {\theta\over 2} $ stands for the
momentum transfer.

\bigskip

The variational expectation value of $ \cos \left[ \beta \phi(x) \right] $
[eq.(\ref{ansatzsingor})] depends on the coordinate $x$ indicating that the
variational soliton state is not a momentum eigenstate but a
superposition of momentum eigenstates,
$$
|\Psi> = \int dp \; f(p) \; |\, p=M\sinh \theta >
$$
where $ | p > $ stands for quantum one-soliton states and $ f(p) $ is
some function normalized such that
\begin{equation}\label{norf}
\int | f(p) |^2 dp = 1 \; .
\end{equation}

Therefore,
\begin{eqnarray}\label{fforma}
<\Psi | T_{\mu}^{\mu}(x)  |\Psi> &=& \int_{-\infty}^{+\infty} dp \, dp'\;  
f(p)^* f(p') \;
e^{i(p-p')x} \;   <p | T_{\mu}^{\mu}(0)  |p'>\cr \cr
 &=&\int_{-\infty}^{+\infty}  \;dp \, dp'\;
F(\theta(p,p')) \; e^{i(p-p')x }\; f(p') \; f(p )^*
\end{eqnarray}
where $ F(\theta) $ is the form factor (\ref{ff})
and $ M^2\, \cosh \theta =
\sqrt{M^2 + p'^2} \sqrt{M^2 + p^2}- p p' $.

Taking the Fourier transform of eq.(\ref{fforma}) yields,
\begin{eqnarray}\label{ffexa2}
\int_{-\infty}^{+\infty} {dx \over 2 \pi} \;e^{ix p } \;
<\Psi | T_{\mu}^{\mu}(x)  |\Psi> &=&  \int dq \
 \; f(q)^* \; f(q + p ) \; F(\theta_{p,q}) \cr \cr
\mbox{where}\; \theta_{p,q}&=&Argsinh{q+p\over M} - Argsinh{p\over M}
\end{eqnarray}

We read  from eq.(\ref{ansatzsingor}) that
$$
<\Psi | T_{\mu}^{\mu}(x)  |\Psi>_{var} = \frac2{\beta^2} \left(
M^2- \frac{2m^2}{\cosh^2[mx]} \right)\; .
$$
Therefore,
\begin{eqnarray}\label{ffvar}
\int_{-\infty}^{+\infty} {dx \over 2 \pi} \;e^{ix p } \;
<\Psi | T_{\mu}^{\mu}(x)  |\Psi>_{var}=
- { 2 p\over {\beta^2 \sinh\left[{\pi p \over 2 m}\right]}} \;.
\end{eqnarray}

We can match the exact expressions and the variational approximation
in the weak coupling  limit which is also a
non-relativistic limit $ p, p' << M, \; \theta <<1 $ since the solitons
becomes very massive for small $ \beta $. Then, $
\theta_{p,q} \approx {q\over M} $ and
$$
\int_{-\infty}^{+\infty} {dx \over 2 \pi} \;e^{ix p } \;<\Psi |
T_{\mu}^{\mu}(x)  |\Psi> \buildrel {p \to 0} \over=
\int_{-\infty}^{+\infty} {dx \over 2
\pi} \;e^{ix p } \; <\Psi | T_{\mu}^{\mu}(x)  |\Psi>_{var}\buildrel {p
\to 0} \over=  -{M \over 2 \pi}
$$
where we used eqs.(\ref{ffexa}), (\ref{norf}) and (\ref{ffexa2}) and
that $ M = 8 m / \beta^2 $  for the  soliton mass in the classical $
\beta \to 0 $ limit.

We do not see for large $ \theta $ Regge behaviour in
$ F_{var}(\theta) $. Eq.(\ref{ffvar}) suggests that $ F_{var}(\theta)
$ decreases exponentially for large momentum transfer.

\bigskip

In summary, the simple self-consistent ansatz (\ref{ansatzsingor}) does not  reproduce the exact on-shell sine-Gordon except in the weak coupling
regime. 


Thus this initial configuration will undergo non-trivial time
evolution to relax  via meson emission to a stationary state of lowest
energy compatible with the variational ansatz.

 Since
for weak coupling the variational initial configuration does not differ
much from the exact solution, we expect slow time evolution at weak
coupling with little meson production. Clearly a deeper insight into
the further evolution of these variational initial configurations
require a numerical study of eqs.(\ref{selfsingor})-(\ref{flucsingord}) [see  
sec. VII].

\subsection{ $\varphi^4 $ Field Theory }

The $\varphi^4$ quantum field theory is the simplest theory that
presents a symmetry breaking phase transition with relevant low-energy
phenomenology in $3+1$ dimensions. It is also a model to describe
kink-like excitations and the collective behavior of quasi-one
dimensional charge density wave systems within the Landau-Ginzburg
framework.
The kink configurations in the broken symmetry phase are interpreted
as domain walls between coexisting degenerate phases. We will study
the $3+1$ dimensional version in the form of the linear sigma model in
the next section. In this section we study the simpler $1+1$ dimensional
model not only as a preliminary step towards the $3+1$ dimensional
implementation but also because this is relevant in statistical
mechanics and condensed matter to describe {\em non-perturbatively}
the quantum
dynamics of transport in quasi-one-dimensional charge density wave
systems\cite{schrieffer,yu,gruner}.

The Lagrangian density is given by
\begin{equation}
{\cal L} = \frac{1}{2} \partial_{\mu}\varphi \partial^{\mu}\varphi
+\frac{\mu^2}{2} \varphi^2 - \frac{g}{4} \varphi^4 \label{fi4lag}
\end{equation}
Upon shifting the field by the expectation value $\varphi_c$, the
Hartree factorization advocated in the previous section leads to the
following equations of motion
\begin{eqnarray}
&& \left[{\partial^2  \over \partial t^2} - {\partial^2  \over \partial x^2}  
- \mu^2 + g \varphi^2_c
+3G(x,x;t)\right]\varphi_c(x,t) = 0  \label{clasfi4} \\
&& \left[{\partial^2  \over \partial t^2} - {\partial^2  \over \partial x^2}  
- \mu^2 +
3g\varphi^2_c+3G(x,x;t)\right]\hat{\psi}(x,t) = 0  \label{heisfi4} \\
&& G(x,x;t) = \langle \hat{\psi}^2(x,t)\rangle \label{selfconfi4}
\end{eqnarray}

\subsubsection{\bf The vacuum sector and the phase transition}
Before proceeding to set up the initial value problem for inhomogeneous
time dependent configurations we will address the properties of the
vacuum sector in the variational approximation. For this we look
at space-time independent configurations of $\varphi_c$ and obtain
the effective potential $V_{eff}(\varphi_c)= \langle {\cal H} \rangle$
with ${\cal H}$ the Hamiltonian density as introduced in section II.
For this purpose it proves convenient to introduce the composite operator
\begin{equation}
\hat{\chi}(x,t) = -\mu^2 +3 g \hat{\varphi}^2(x,t)\label{chiop}
\end{equation}
whose expectation value in the Gaussian variational state is given by
\begin{equation}
\chi(x,t) = -\mu^2+3g\varphi^2_c(x,t)+3gG(x,x;t)\label{chiexp}
\end{equation}
which is recognized as the space-time dependent effective mass of the
quantum fluctuations.

For a space time constant configuration corresponding to the Fock
vacuum of the annihilation operators, we find
\begin{equation}
G(x,x) = \int \frac{dk}{2\pi} \frac{1}{2\sqrt{k^2+\chi}}
\end{equation}
and leads to the self-consistent gap equation
\begin{equation}
\chi = -\mu^2+3g\varphi^2_c+3g \int \frac{dk}{2\pi}  
\frac{1}{2\sqrt{k^2+\chi}} \label{gap}
\end{equation}
Since the Heisenberg equations (\ref{clasfi4},\ref{heisfi4}) must
be finite, the logarithmic divergence in the integral is absorbed into
a mass renormalization. This polynomial theory is superrenormalizable
in $1+1$ dimensions and mass renormalization renders the theory finite.

The effective potential  is given by
\begin{equation}
V_{eff}  =  < V > = -{1 \over 2} \mu^2  \phi^2 + {g \over 4} \phi^4 -
{\mu^2 G \over 2}
+ {3 g \over 4} G^2 + {3 g \over 2} G \phi^2 \label{veff}
\end{equation}
which can be written in terms of the two variables $\varphi_c \; ; \;
\chi$ (the composite operator effective potential\cite{jackiw}) in the form
\begin{equation}
V_{eff}(\phi,\chi) = {1 \over 2} \chi \phi^2  - {g \over 2}
\phi^4 - {1 \over 6 g} [{\chi^2 \over 2} + \mu^2 \chi] + {1 \over 2} \int {dk  
\over 2 \pi} {1 \over 2 \omega(k)}
\end{equation}
This form of the effective potential is useful because
$\partial V_{eff}(\varphi_c,\chi)/\partial \chi$ gives the gap equation
(\ref{gap}).

As a function of two variables we must minimize the potential with
respect to  both to find the vacuum state.
The minimum of the potential is obtained from the condition
\begin{equation}
\chi \phi - 2 g \phi^3 = 0.
\end{equation}
The value of $\phi$ at the minimum is the vacuum expectation value $v$ and
the expectation value of the composite operator evaluated at this minimum is
given by
\begin{equation}
  m_r^2 \equiv \chi_{\phi=v} = 2 g v^2. \label{chimin}
\end{equation}
The gap equation at the minimum is given by
\begin{equation}
 2 g v^2  = -\mu^2 + 3 g \; \left[v^2 + \int {dk \over 2 \pi}
{1 \over 2 \omega_0(k)} \right]
\end{equation}
where $\omega_0(k) = (k^2+m_r^2)^{1/2}$. We can use this value to
subtract the gap equation for arbitrary $\varphi_c$, and the renormalized gap  
equation then becomes:
\begin{equation}
\chi= 2g v^2 + 3 g (\phi^2 - v^2) -{3 g \over 4 \pi} \log[{\chi \over m_r^2}]
\end{equation}

Integrating the renormalized expression for ${\partial V \over \partial
\chi}$ with respect to $\chi$ gives us  the renormalized effective potential  
which is:
\begin{eqnarray}
V[\phi,\chi] &&= {\chi \over 2} (\phi^2 -v^2) -{g \over 2} (\phi^4 - v^4)
\nonumber \\
&& -{1 \over 12 g} (\chi -m_r^2)^2 + {1 \over 8 \pi} \left[ \chi -m_r^2 +
\chi  \log ({m_r^2 \over \chi})\right]\; .
\end{eqnarray}
and we have chosen the potential to be zero at the minimum for convenience.
As a function of the coupling $g$ this effective potential describes
a (weak) first order phase transition as shown explicitly below.

It is convenient to now rescale everything in terms of the renormalized mass  
$m_r^2 =  2 g v^2$. we let
\[ g \rightarrow m_r^2 g;, \chi \rightarrow m_r^2 \chi \]
so that at the minimum we now have:
\[ \chi=1;  ~~ \phi^2 ={1 \over 2g} \]
The effective potential is now given by:
\begin{eqnarray}
{V[\phi,\chi,g] \over m_r^2} &&= {\chi \over 2} (\phi^2 -{1 \over 2 g}) -{g
\over 2} (\phi^4 - {1 \over 4 g^2}) \nonumber \\
&& -{1 \over 12 g} (\chi -1)^2 + {1 \over 8 \pi} ( \chi -1 - \chi \log \chi).
\end{eqnarray}
This can be most easily plotted by using the relationship:
\[ \phi^2 = {1 \over 2 g} + {1 \over 3 g} (\chi -1) + {1 \over 4 \pi} \log \chi \]
and parametrically plotting $V$ and $\phi$ as a function of $\chi$.
To find the position of the phase transition, this takes place when the value
of the potential at the
minimum at $\phi=0$ has the same value as at the minimum at $\phi=v$ which
we chose to be zero.
Solving for this numerically we find that
\begin{equation}
g_c/m_r^2 = .3896 \pi = 1.224.
\end{equation}
To show this behavior we plot the effective potential at the three values
$g/m_r^2 = 1, g_c, 2$ in Fig. 1.

For $g<g_c$ the potential has a double well structure with degenerate
minima at $\varphi_c = \pm v$, as the value of the (dimensionless)
coupling is increased the minima shifts abrubtly to $\phi=0$ when $ g
\geq g_c$, these results are similar to those found
previously\cite{tdhf1}.

\subsubsection{\bf Inhomogeneous Initial value problem: Kinks and Lumps}
The harmonic fluctuations around a stationary classical kink
configuration are solutions of a Schr\"odinger-like eigenvalue problem
with a potential given by eq. (\ref{integrable}) for $ n=2 $ whose
solutions
are also known exactly\cite{morse,rajaraman}. The classical kink
configuration corresponds to
\begin{equation}
\varphi^2_c(x) = v^2 \left[1-\frac{1}{\cosh^2[mx]}\right] \label{ficlas}
\end{equation}
with $v$ the tree level vacuum expectation value. Thus we seek a
self-consistent solution of the eigenvalue problem in the Gaussian
variational approximation of the form
\begin{equation}
\varphi^2_c(x)  =  v^2-\frac{\alpha}{\cosh^2[mx]}
\; \; ; \; \; \langle \hat{\psi}^2(x) \rangle = \langle
\hat{\psi}^2(\infty) \rangle + \frac{A}{\cosh^2[mx]} \label{selfish}
\end{equation}
We propose a variational ansatz similar to that of the Sine
Gordon problem above
\begin{equation}
-\mu^2+3g\left[\varphi^2_c+G(x,x)\right] = M^2 -\frac{2m^2}{\cosh^2[mx]}
\label{ansatzfi4}
\end{equation}
where the variational parameters $M\; ; \; m$ will be determined
self-consistently. The logarithmic divergence in $<\psi^2(x)>$ is
independent of $x$ and is absorbed in mass renormalization, which
is the {\em only} renormalization required in this superrenormalizable
theory. Thus define the renormalized mass parameter
\begin{equation}
\mu^2_R = \mu^2 - 3gG_0 \; \; ; \; \; G_0 = \langle \hat{\psi}^2(\infty)  
\rangle \label{renomassfi4}
\end{equation}
Requiring that $\varphi_c(x)$ asymptotically for large $|x|$ becomes an {\em  
exact} solution of the space-time dependent problem leads to the
constraint
\begin{equation}
-\mu^2_R v+ gv^3 =0 \; \Rightarrow v^2 = \frac{\mu^2_R}{g} \label{vevfi4}
\end{equation}
Then the asymptotic solution of the eigenvalue problem leads to the
identification
\begin{equation}
M^2 = 2 \mu^2_R \label{massrenfi4}
\end{equation}
which is the usual relationship between the mass of the asymptotic
quanta and the vacuum expectation value in the broken symmetry phase.

Since the ansatz (\ref{ansatzfi4}) is similar to that in the Sine Gordon case  
studied previously, the results for $G(x,x)$ can be obtained from
the previous case and is given by eq.(\ref{GofxSG}) with
\begin{equation}
G_0 = \int \frac{dk}{2\pi} \frac{1}{2\sqrt{k^2+M^2}} \label{G0fi4}
\end{equation}
Finally the expressions (\ref{GofxSG},\ref{G0fi4}) with
the identifications given by  
(\ref{renomassfi4},\ref{vevfi4},\ref{massrenfi4}) and the ansatz for the  
expectation value (\ref{ficlas}) lead to the self-consistent
solution
\begin{equation}
\alpha = \frac{2m^2}{3g}+F[\eta] \label{alfafi4}
\end{equation}
with $F[\eta]$ given in (\ref{GofxSG},\ref{F}). The only possibility for
a topological kink profile is given by
\begin{equation}
\alpha = v^2 \Rightarrow \varphi^2_c(x) = v^2 \tanh^2[mx] \label{topofi4}
\end{equation}

Requiring that the initial configuration $\varphi_c(x)$ have a topological
kink profile, the self-consistency
condition (\ref{alfafi4}) leads to  the following transcendental equation
for $\eta$
\begin{equation}
\frac{3-\eta^2}{3(1+\eta^2)F[\eta]} = \frac{g}{\mu^2_R} \label{xieqnfi4}
\end{equation}
where we have used eqs.(\ref{topofi4},\ref{vevfi4},\ref{massrenfi4}) and
the expression for $\eta$ given in eqs. (\ref{GofxSG},\ref{F}), the
right hand side of (\ref{xieqnfi4}) is recognized as the dimensionless
coupling. It is clear that there are solutions for {\em any} value
of the dimensionless coupling with $0<\eta < 3$.

We note that this variational form for $\varphi_c(x)$ is {\em not} a stationary
solution of the equation of motion (\ref{clasfi4}) but must be taken as
an {\em initial} configuration that will eventually evolve in time.

This method also allows us to construct non-topological lump solutions
with $\alpha < v^2$. These solutions correspond to initial inhomogeneous
configurations that will relax via emission of mesons.

Although we have renormalized the theory at the initial time by mass
renormalization, it is straightforward to prove that the same mass
renormalization can be performed at all times. The main point is that
the Heisenberg equations for the operator $\psi$ can be written in the
form
\begin{equation}
\left[{\partial^2  \over \partial t^2} - \frac{\partial^2}{\partial
x^2} + M^2 + \delta M^2(x,t) 
\right]\psi(x,t)=0 \label{1dheispert}
\end{equation}
and the mass counterterm $\delta M^2(x,t)$ is a localized function
of space. The Green's function can be calculated in perturbation theory
in $\delta M^2(x,t)$ in terms of Feynman diagrams with free field
propagators of mass $M$ and mass insertions of $\delta M^2(x,t)$. The
zeroth order has a logarithmic divergence which is absorbed in mass
renormalization and all higher order insertions are finite as long as
the theory remains superrenormalizable, which in the case of Sine-Gordon
requires $\beta^2 < 8\pi$. This is equivalent to obtaining the
divergences in an operator product expansion.

\subsection{\bf Variational approach vs. Collective Coordinates:}
A perturbative approach to quantization around a soliton configuration
begins by obtaining the eigenmodes of the Schr\"odinger operator
associated with the harmonic fluctuations around the classical
kink\cite{rajaraman,lee}-\cite{devega}. There is an eigenmode
with zero frequency associated with translational
invariance\cite{rajaraman,lee}-\cite{devega} that must be
treated non-perturbatively. This mode is isolated from the mode
expansion and
treated as a collective coordinate\cite{rajaraman,lee}-\cite{devega}
(although see\cite{fad} for a different approach) and perturbation
theory is carried out for the non-zero frequency modes, since these
can be studied in a small amplitude expansion. Collective coordinate
quantization is complicated by the fact that the functional basis is
not cartesian\cite{lee,c3,c4} with complicated Jacobians and a
prescription for normal ordering\cite{c1,c2,c5,devega}. Although high
order calculations of S-matrix elements\cite{c1,c2,c5,devega} and more
recently of non-equilibrium properties\cite{boysoli} had been carried
out, they represent an enormous technical feat.

In order to make contact with the collective coordinate program of
quantization and the treatment of zero modes, we recognize that
in the self-consistent solution for the fluctuations, the mode that
would correspond to the zero frequency mode in the perturbative approach
is precisely $\psi_0(x)$ with frequency $\omega_0=\sqrt{M^2-m^2}$ (see
\ref{zeromodeSG}), that indeed becomes zero for $m=M$. However this value of  
$m$ leads to infrared divergences in our treatment (the variable $\eta$ given  
by eq. (\ref{GofxSG}) diverges), which are a reflection of the infrared  
divergences that appear in the perturbative approach.  We notice, however,  
that self-consistent solutions exist for a large range
of the ratio $m/M$ as discussed above and there are no infrared pathologies  
associated with any value of $m$ for which solutions are available.

The apparent contradiction between the variational approach and the  
collective coordinate method is resolved when it is realized that i) the
profile $\varphi_c(x)$ obtained from the self-consistent solution to
the mode equations {\em is not} a stationary solution of the equation
of motion including the quantum backreaction effects. ii) the zero mode
in the perturbative treatment is a consequence of translational
invariance, the change in the classical field profile under an
infinitesimal shift of the origin is the zero mode of the fluctuation
operator\cite{rajaraman,jackiw,lee}-\cite{devega}. In the variational
approach translational invariance is {\em also} present since the
ansatz for the self-consistent potentials
(\ref{ansatzsingor},\ref{ansatzfi4}) could be written in terms of
$x-x_0$ with $x_0$ an arbitrary point (the center of the soliton or
lump) and the mode functions will be functions of $x-x_0$. However,
unlike in the
perturbative calculation in which an infinitesimal translation of the
center of the classical configuration leads to a zero mode whose
wave-function is $\propto d\varphi_c(x)/dx$, in the self-consistent
approach, such an invariance requires an infinitesimal translation of
$\varphi_c$
{\em and the self-consistent potential} which involves
$G(x,x)$. Obviously the function obtained by an infinitesimal shift in
$x_0$ from
$\varphi_c$ is {\em not} a zero mode of the  fluctuation operator,
precisely because of the self-consistent potential that provides the
quantum backreaction.

Thus our interpretation of the resolution of the apparent
contradiction is that the quantum backreaction effects that dress
the soliton or
lump solution also remove the degeneracy associated with the zero mode.
Although translational invariance is present, the self-consistent ansatz
determines a profile centered at a particular point in space, shifting
the center is tantamount to shifting the spatial dependence of all of the
Green's functions and expectation values under which the expectation value of 
the Hamiltonian in the trial state is invariant.

Thus this self-consistent variational method based on a trial ansatz,
provides a non-perturbative manner of circumventing the infrared
singularities associated with translational zero modes in the perturbative
approach.

\section{\bf $3+1$ Dimensions: Inhomogeneous Chiral Condensates}

One of our main motivations for studying inhomogeneous non-equilibrium
problems is to provide a consistent framework to study the relaxation
of high energy inhomogeneous configuration in heavy ion collisions via
expansion into the vacuum. The main goal is to provide a microscopic
non-equilibrium description of the dynamics leading to a hydrodynamic
description and the scaling solutions. Of particular relevance is
the study of the relaxation of  inhomogeneous chiral condensates since
this could be potential soft probes of the chiral phase transition
in the form of enhanced pion distributions with distinct isospin
correlations. In order to address these issues we must i) provide
a model quantum field theory that incorporates the relevant low
energy physics, ii) we must propose an initial state that incorporates the
relevant geometry, leading to longitudinal or spherical expansion,
iii) provide a consistent framework to study the microscopic dynamics.

For the first step we invoke the linear model in terms of a sigma field
and an isospin triplet of pseudoscalar pions. In order to maintain the
phenomenology of PCAC in the Gaussian approximation we invoke the large
$N$ limit in which the sigma field and the pions are in the vector
representation of the $O(N)$ group. For the initial inhomogeneous field
configuration one should ideally propose either an inhomogeneous
cylindrically symmetric field configuration in the sigma component (or
possibly with an inhomogeneous component along a definite isospin direction
if one wants to study a disoriented configuration) reminiscent of two
pancakes or a disk configuration with an extension
in the transverse direction of several $\mbox{fm}$.

The study of such a configuration including quantum effects as advocated in
this work is an extremely complicated task and beyond the computational
capabilities at this moment. We will be less ambitious and
seek to understand the dynamics of a simpler configuration with spatial
variation along only one coordinate, chosen to be the z-axis and
translationally invariant in the perpendicular directions.
Even this simpler scenario represents an advance in the treatment of
semiclassical configurations {\em plus backreaction}.
 Lastly, the
time evolution is to be studied with the methods detailed in the previous
section. As in the simpler $1+1$ dimensional problems we begin by
establishing the self-consistent initial value problem and Cauchy data
on the necessary Green's functions and inhomogeneous configurations.
We note that rather than invoking boost invariance that would imply
specifying Cauchy data on a proper time hyper-surface as is the case in
a hydrodynamical treatment, we emphasize that  boost invariant
hydrodynamics must arise from the full microscopic dynamics, and we
set up the Cauchy data on an initial  constant time slice ($t=0$).

The Lagrangian density for the linear sigma model with an explicitly
symmetry breaking term that gives a mass to the pions is given by
\begin{eqnarray}
{\cal{L}} &=&\frac{1}{2}\partial_{\mu}\vec{\Phi}\cdot
\partial^{\mu}\vec{\Phi}-\frac{1}{2}\mu^2_0\;
\vec{\Phi}\cdot\vec{\Phi}+\frac{\lambda}{8N}
(\vec{\Phi}\cdot\vec{\Phi})^2-h\sigma
\label{linsigmamodel} \\
\vec{\Phi} & = & (\sigma,\pi_1,\dots,\pi_{N-1})
\end{eqnarray}
The linear sigma model description of low energy pion phenomenology
and PCAC corresponds to $N=4$ with parameters fit to reproduce the
pion decay constant $F_{\pi} = 93\mbox{Mev}$, the pion mass
$ M_{\pi} \approx 140\mbox{Mev}$ and the sigma mass (the value of
$\sqrt{s}$ at
which the phase shift in the pseudoscalar channel in $\pi-\pi$ scattering
goes through $\pi/2$) $M_{\sigma} \approx 600\mbox{Mev}$.

In order to implement the Gaussian approximation which is appropriate
in the large $ N $ limit we begin by performing the shift along the
sigma direction
\begin{equation}
\sigma (\vec{x},t ) = \varphi_c(\vec x,t)\; \sqrt{N}+
\chi ( \vec{x},t) \; \; ; \; \; \langle \chi(\vec x,t)
\rangle =0 , \label{sigmashift}
\end{equation}
To leading order the large N-limit is implemented by considering a Hartree-like
factorization (neglecting 1/N terms)
\begin{eqnarray}
\chi^4 & \rightarrow & 6 \langle \chi^2 \rangle \chi^2 +\mbox{ constant },
\label{larg1} \\
 \chi^3 & \rightarrow & 3 \langle \chi^2 \rangle \chi,
\label{larg2} \\
 \left( \vec{\pi} \cdot \vec{\pi} \right)^2 & \rightarrow &
2 \langle \vec{\pi}^2 \rangle \vec{\pi}^2 - \langle \vec{\pi}^2 \rangle^2+
{\cal{O}}(1/N), \label{larg3} \\
\vec{\pi}^2 \chi^2 & \rightarrow & \langle
\vec{\pi}^2 \rangle \chi^2 +\vec{\pi}^2 \langle \chi^2 \rangle,
\label{larg4} \\
 \vec{\pi}^2 \chi & \rightarrow & \langle \vec{\pi}^2
\rangle \chi.  \label{larg5}
\end{eqnarray}

 and assuming $O(N-1)$ invariance by
writing

\begin{equation}
\vec{\pi}(\vec x, t) = \psi(\vec x, t)
\overbrace{\left(1,1,\cdots,1\right)}^{N-1} \Rightarrow
\langle \vec{\pi}\cdot \vec{\pi} \rangle = N \langle \psi^2 \rangle
\label{oninva}
\end{equation}

Alternatively the large $ N $ expansion is systematically implemented by
introducing an auxiliary field\cite{losalamos,largeN}, this latter procedure
allows a consistent implementation beyond the leading order.  To leading
order the two methods are equivalent\cite{erice97}. This factorization can
also be understood by implementing
the tadpole method described in section II. 


Thus to leading order, the large N approximation is similar to the
Hartree variational approach described in section II. The Hartree and large N equations of motion
are similar but not identical,
they differ by a factor of three between the classical and quantum contributions\cite{noneq}. 

The tadpole equation
leads to the evolution equation for the expectation value
\begin{equation}
\left[{\partial^2  \over \partial t^2}-\nabla^2 -\mu^2_0 +
\frac{\lambda}{2}\varphi^2_c(\vec x,t)+ \frac{\lambda}{2}\langle
\psi^2(\vec x,t) \rangle \right]\varphi_c(\vec x,t) = h
\label{expeclargen}
\end{equation}
which is the mean-field equation obtained via the auxiliary field
method\cite{losalamos,largeN}.
After the large N-Hartree factorization and the implementation of the
tadpole condition, the Hamiltonian for the field $\psi$ becomes
quadratic and
the Heisenberg equation for this field becomes

\begin{eqnarray}
&& \left[{\partial^2  \over \partial t^2}-\nabla^2 +M^2(\vec x,t) \right]\psi(\vec
x,t) = 0 \nonumber \\
&& M^2(\vec x,t) =
-\mu^2_0 + \frac{\lambda}{2} \varphi^2_c(\vec x,t)+ \frac{\lambda}{2}
\langle \psi^2(\vec x,t) \rangle  \label{flucslargen}
\end{eqnarray}

We will propose a configuration $\varphi_c$ that only depends on one
spatial coordinate and time, i.e. $\varphi_c(z,t)$. This simplification
will allow us to make progress in setting up the initial value problem
and to make use of the methods
developed for the $1+1$ dimensional case. Thus for the initial value
problem we will solve the eigenvalue problem (\ref{modeeqn}) with the
ansatz
\begin{eqnarray}
\varphi^2_c(z)  & = &  F^2_{\pi} +\frac{\alpha}{\cosh^2[mz]} \label{expec3d} \\
M^2(\vec x) & = & -\mu^2_0 + \frac{\lambda}{2} \varphi^2_c(\vec x,t=0)+  
\frac{\lambda}{2} \langle \psi^2(\vec x,t=0) \rangle
= M^2_{\pi}-\frac{2m^2}{\cosh^2[mz]} \label{flucs3d}\\
\end{eqnarray}

Asymptotically the solutions should be phase shifted plane waves describing
asymptotic pions quanta of mass $M_{\pi}$ and therefore asymptotically for $x  
>> 1/m$ the
PCAC relation $M^2_{\pi}F_{\pi}=h$ is fullfiled with
\begin{equation}
M^2_{\pi}=-\mu^2_0 + \frac{\lambda}{2}F^2_{\pi}+ \frac{\lambda}{2} \langle  
\psi^2(\infty) \rangle \label{pionmass}
\end{equation}
is the {\em renormalized} pion mass. Translational invariance in the
perpendicular directions allows us to write the mode functions solutions
of the eigenvalue problem as
\begin{eqnarray}
&& \psi_k(\vec x) = e^{i\vec{k}_{\bot}\cdot \vec{x}_{\bot}}\psi_k(z)  
\label{planewaves} \\
&& \left[\frac{d^2}{dz^2}+W^2_k -m^2 +   
\frac{2m^2}{\cosh[mz]}\right]\psi_k(z)=0 \label{eigeneq3d} \\
&&W^2_k = \omega^2_k - \vec{k}^2_{\bot}-M^2_{\pi}+m^2 \label{bigfreq}
\end{eqnarray}

We will take as initial state a Gaussian wavefunctional corresponding
to the ground state of the harmonic oscillators with frequencies $\omega_k$.

In terms of the frequencies $W^2_k$ the problem is now exactly the
same as the $1+1$ dimensional problems that were solved self-consistently
in the previous section leading to the eigenfunctions and frequencies given by
\begin{eqnarray}
\psi_0(\vec x) & = & e^{i\vec{k}_{\bot}\cdot \vec{x}_{\bot}}  
\sqrt{\frac{m}{2}} \frac{1}{\cosh[mz]} \; \; ; \; \; \omega^0_k =
\sqrt{\vec{k}^2_{\bot}+M^2_{\pi}-m^2} \label{zeromode3d} \\
\psi_k(\vec x) & = &
 \frac{e^{i\vec{k}\cdot \vec{x}}}{\sqrt{k^2_z+m^2}}
\left[ik_z -m\tanh[mz]\right] \; \; ; \;
\omega_k= \sqrt{\vec{k}^2+M^2_{\pi}} \label{kmodes3d}
\end{eqnarray}

\noindent where again stability requires that $M^2_{\pi} > m^2$.
In the perturbative quantization approach, the transverse modes associated
with $\psi_0(\vec x)$ are the  `capillary' waves associated with massless
fluctuations transverse to the interface with dispersion
relation $\omega^0_k= |\vec k|$ as a consequence of the translational
zero mode in the z-direction.

 At the initial
time the Heisenberg field operator is expanded in terms of this complete
set of mode functions. With a continuum normalization

\begin{eqnarray}
\psi(\vec x) & = &  \int \frac{d^3k}{(2\pi)^{3/2}}
\frac{1}{\sqrt{2\omega_k}} \left\{
a(\vec k) \psi_k(\vec x) + h.c. \right\} + \int
\frac{d^2\vec{k}_{\bot}}{(2\pi)}
\frac{1}{\sqrt{2\omega^0_k}} \left\{ a_0(\vec k)\psi_0(\vec x)+
h.c. \right\} \label{modexp3d}\\
\pi(\vec x) & = &  -i\int \frac{d^3k}{(2\pi)^{3/2}}
\sqrt{\frac{\omega_k}{2}} \left\{
a(\vec k) \psi_k(\vec x) - h.c. \right\} -i
\int \frac{d^2\vec{k}_{\bot}}{(2\pi)}
\sqrt{\frac{\omega^0_k}{2}} \left\{ a_0(\vec k)\psi_0(\vec x)-
h.c. \right\} \label{pimodexp3d}
\end{eqnarray}

Taking the initial state to be the vacuum for all the annihilation
operators we finally find
\begin{eqnarray}
&&\langle \psi^2(\vec x) \rangle  =   \langle \psi^2(\infty) \rangle +
I \nonumber \\
&&\langle \psi^2(\infty) \rangle  =  \int \frac{d^3k}{(2\pi)^3}
\frac{1}{2\sqrt{\vec{k}^2+M^2_{\pi}}} \label{psiinfty}\\
&&I  =
\frac{m^2}{\cosh^2[mz]} \left[ \frac{1}{2m} \int
\frac{d^2\vec{k}_{\bot}}{(2\pi)^2}
\frac{1}{2\sqrt{\vec{k}^2_{\bot}+M^2_{\pi}-m^2}}- \int
\frac{d^3k}{(2\pi)^3} \frac{1}{2\sqrt{\vec{k}^2+M^2_{\pi}}} \;
\frac{1}{k^2_z+m^2} \right]
\label{psi23d}
\end{eqnarray}

The first term is the usual coincidence limit of the two point function
and has a quadratic and logarithmic divergence which as usual is
absorbed in a mass renormalization. Introducing an ultraviolet cutoff
the  integrals in I  yield to
\begin{eqnarray}
I & = &  \frac{m^2}{\cosh^2[mz]}\left\{ \frac{1}{4\pi^2}
\ln\left[\frac{\Lambda}{M_{\pi}} \right] +
H\left[\frac{M_{\pi}}{m}\right]\right\} = \frac{m^2}{\cosh^2[mz]}\;
\tilde{I} \nonumber\\
H[x]  & = &  \frac{1}{4\pi^2} \left\{1-\sqrt{x^2-1} \;
\mbox{ArcTg}\frac{1}{ \sqrt{x^2-1}} \right\}
\label{hfunc3d}
\end{eqnarray}
Finally the ansatz (\ref{expec3d},\ref{flucs3d}) leads to the following  
self-consistent condition for the effective mass term in the mode equation
\begin{equation}
-\frac{2m^2}{\cosh^2[mz]}= \frac{\lambda  
m^2}{2\cosh^2[mz]}\left[\alpha+\tilde{I}\right] \label{finanzatz}
\end{equation}

The logarithmic divergence is absorbed in coupling constant renormalization
\begin{equation}
\frac{1}{\lambda_R} =  
\frac{1}{\lambda}+\frac{1}{16\pi^2}\ln\left[\frac{\Lambda}{M_{\pi}}\right]
\label{renorcoup}
\end{equation}
where we chose to renormalize at the scale $M_{\pi}$. After renormalization,  
the self-consistent solution for the inhomogeneous expectation value becomes
\begin{equation}
\varphi^2_c(z) =  
F^2_{\pi}\left[1-\left(\frac{4}{\lambda_R}+H\left[\frac{M_{\pi}}{m}\right]\right)\frac{m^2}{F^2_{\pi}\cosh^2[mz]}\right]  
\label{selfcons3d}
\end{equation}
This self-consistent solution for
the initial value problem is {\em not} a stationary solution of the
evolution equation for the expectation value. This field profile will
evolve in time and dissipate into meson excitations. The dynamics of
this process of relaxation is studied within the time dependent scheme
described above and addressed in detail below.

The mass renormalization and coupling constant renormalization make finite the
effective space dependent mass term $M^2(\vec x,t=0)$, we now show that this  
renormalization
renders the effective  mass term finite self-consistently at {\em all times}.  
The proof is an implementation
of the operator product expansion and goes as follows. Since the  
inhomogeneity is localized
in space and  asymptotically the expectation value $\varphi_c(\vec x,t)$  
reaches its vacuum
value, we can write
\begin{equation}
M^2(\vec x,t)  =   M^2_{\pi}+ \delta M^2(\vec x,t) \label{masspert}
\end{equation}
\noindent The renormalization is obtained by requiring that the equations of motion
are finite which results in the requirement that the effective mass term
$M^2(\vec x,t)$ be finite for all values of the space time arguments. Since  
$M^2_{\pi}$ is the renormalized pion mass, renormalizability implies that

\begin{equation}
\delta M^2(\vec x,t) = \frac{\lambda}{2}\left[ G(\vec x,\vec x,  
t)-G_0+\varphi^2_c(\vec x,t)-
\varphi^2_c(\infty) \right] = \frac{\lambda_R}{2}\left[ G_R(\vec
x,\vec x, t)+\varphi^2_c(\vec x,t)-
\varphi^2_c(\infty) \right]
\label{deltamass}
\end{equation}

\noindent is finite.

The divergences in $G(\vec x,\vec x,t)$ can be obtained by implementing the
operator product expansion or alternatively by treating the  term
$\delta M^2(\vec x,t)$  as a localized perturbation, i.e. as a
mass insertion in the  Green's function.

It is a straightforward exercise to expand the
Green's function $G(\vec x,\vec x,t)$ which is the coincidence limit
of the Feynman propagator in a perturbative Feynman series. This is a
tadpole diagram (a two-point function in the coincidence limit) in
the presence of a background mass insertion, the zeroth order (one
loop) is simply $G_0 = \langle \psi^2(\infty) \rangle$ given by
eq. (\ref{psiinfty}) and which has been absorbed in the mass
renormalization and is cancelled by the regular (space-time
independent) mass counterterm $G_0$ in (\ref{deltamass}). The first
order in $\delta M$ to one loop is simply a mass
insertion in the tadpole that has exactly the divergence of the
one-loop contribution to coupling constant renormalization (actually
the renormalization of the composite operator $<\psi^2>$ which is
equivalent to coupling renormalization in the large $ N $ limit). Since
the divergence is independent of the momentum transfer into the loop
and higher order insertions are finite by power counting   we find the
result
\begin{equation}
G(\vec x,\vec x, t)-G_0 =
-\frac{1}{8\pi^2}\ln\left[\frac{\Lambda}{M_{\pi}}\right]\;
\delta M^2(\vec x,t) +
\tilde{G}_{R}(\vec x,\vec x,t)
\end{equation}
with $\tilde{G}_{R}(\vec x,\vec x,t)$ being finite. This expression
leads at once to the
renormalization of coupling and mass just as at $t=0$. Writing
\begin{eqnarray}
\lambda & = &  Z_{\lambda} \lambda_R  \label{barecoup} \\
 Z_{\lambda} & = & \left[
 1-\frac{\lambda_R}{16\pi^2}\ln\left[\frac{\Lambda}{M_{\pi}}\right]
 \right]^{-1}
\label{zlambda}
\end{eqnarray}
then the quantity
\begin{equation}
\delta M^2(\vec x,t) =  \frac{\lambda_R}{2}Z_{\lambda}\left[ G(\vec
x,\vec x, t)-G_0+\varphi^2_c(\vec x,t)-
\varphi^2_c(\infty) \right] \label{finitemass}
\end{equation}
\noindent is {\em finite}.  A more elegant and systematic treatment of
 the renormalization along the lines described above can be offered
 following the analysis presented in reference\cite{baacke}.
 The linear sigma model description of low energy pion phenomenology
dictates that in order to reproduce the values of $F_{\pi}\approx
 93\mbox{Mev},\; M_{\pi}  \approx 140\mbox{Mev} $ and the sigma
 mass $\approx 600 \mbox{Mev}$,
the value of the quartic coupling is extremely large $\lambda_R
 \approx 36$. For such a large value of the coupling there is a Landau
 pole at a value of the cutoff $\Lambda \leq 2-3 \; \mbox{Gev}$
 requiring that this
theory be understood as a low energy theory for scales well below this
cutoff.

In a weakly coupled theory the form of the initial self-consistent
configuration (\ref{selfcons3d}) would allow to prepare initial
inhomogeneous states for which $\varphi_c$
probes the false vacuum $\varphi_c \approx 0$ on length scales much
larger than the Compton
wavelength of the pion. This would be achieved  by specifying a value
$m<<M_{\pi}$ such that $4m^2/\lambda_R F^2_{\pi} \approx 1$. However
in the strongly coupled linear sigma model and in the stable region
for values  of $m < M_{\pi}$ there is no range of the variational
parameter $m$ that would allow to probe the `false vacuum'
region $\varphi_c \approx 0$. The main reason for this constraint is
that we restricted our study to the stable region for which
$M^2_{\pi} > m^2$ to avoid the instabilities associated with the bound
state solutions given by eq. (\ref{zeromode3d}).

\subsection*{\bf Spinodal Instabilities from Klein's paradox}

The nature of the instabilities for $m^2 > M^2_{\pi}$ is clear from
the eigenvalue eq. (\ref{eigeneq3d}) when written in the following form

\begin{equation}
 \left[-\frac{d^2}{dz^2}+\vec{k}^2_{\bot} +M^2_{\pi} -
 \frac{2m^2}{\cosh[mz]}\right]\psi_k(z)= \omega^2_k \psi_k(z)
 \label{spino3d}
\end{equation}

For $ m^2> M^2_{\pi}$ the potential penetrates the negative energy
continuum in a region in space of width $ \Delta z \approx 1/m $ and
there will be a band of perpendicular momenta $k_{\bot}$ for which
there will be instabilities arising from particle production. This
situation is similar to that of the `Klein paradox' for free
quantum fields in the presence of external potentials. In a region
near the origin the potential is negative and strong enough to cross
the threshold $2M_{\pi}$ and produces particles. This instability towards
particle production for the case $m^2 > M^2_{\pi}$ is also encoded in the  
typical square root in the expressions (\ref{hfunc3d}).

For this case there is a band of wavectors $k_{\bot}$ for which there
will be strong production of pions with soft perpendicular momentum.
This mechanism of particle production is very similar to that of
spinodal instabilities in the case of phase transitions, since the modes
of the quantum field feel  an inverted harmonic oscillator potential, i.e. an
effective negative mass squared. We can deal with this interesting situation
in the same manner as in previous treatment of the dynamics as an initial  
value problem in phase transitions with spinodal instabilities.

 The method consists of preparing the initial state wave functional for the  
unstable modes that is the ground state wave functional for upright
harmonic oscillators, with reference frequencies $\tilde{\omega}_k^0=
\sqrt{\vec{k}^2_{\bot}+\mu^2}$, with $\mu^2$ an arbitrary positive mass
parameter. This is the quantum field theory analog of preparing an initial  
Gaussian state, which is the ground state of a particular harmonic oscillator  
potential  and evolving it in an inverted harmonic oscillator potential, which  
is the physical situation for spinodal instabilities.

Therefore we consider the situation with $m^2 > M^2_{\pi}$ by expanding
the field in terms of the eigensolutions $\psi_k$ of (\ref{eigeneq3d})
as in (\ref{modexp3d}) but with the frequencies
\begin{eqnarray}
\Omega_k^0 & = & \sqrt{\vec{k}^2_{\bot}+\mu^2}\quad
\mbox{for} \quad \vec{k}^2_{\bot} < \Delta^2
\label{smallk} \\
\omega_k^0 & = & \sqrt{\vec{k}^2_{\bot}-\Delta^2} \quad \mbox{for} \quad
\vec{k}^2_{\bot} > \Delta^2
\label{largek} \\
\Delta^2 & = & m^2-M^2_{\pi} >0 \label{Delta2}
\end{eqnarray}

and the initial state is annihilated by all of the annihilation
operators. In this case we find the result
\begin{eqnarray}
\langle \psi^2(\vec x,t=0) \rangle &=& \langle \psi^2(\infty)\rangle +
 \tilde{I} \label{newI} \\
 \tilde{I} &=& \frac{1}{4\pi} \ln\left[\frac{\Lambda}{M_{\pi}}\right]
 + F \left[\frac{M_{\pi}}{m},\frac{\mu}{m}\right] \label{itilde}\\
F \left[x,y\right]&=&\frac{1}{8\pi}\left[\sqrt{1+y^2 - x^2} - y\right]
+ \frac{1}{4\pi^2} \left[ 1 + \frac{1}{2}
 \sqrt{1-x^2}\ln\frac{1-\sqrt{1-x^2}}{1+\sqrt{1-x^2}}\right]
\end{eqnarray}
with $\langle \psi^2(\infty)\rangle$ given by (\ref{psiinfty}).

The square roots and the logarithmic branch singularities are the
hallmark of thresholds to particle production and provide an intuitive
explanation for the
spinodal instabilities for large amplitudes: the field profile probes
the negative energy continuum in a localized region in space,
resulting in particle production.

 Renormalization proceeds as in the previous case, and leads to the
final form of the self-consistent field profile for the expectation
value in the case in which $m^2>M^2_{\pi}$

\begin{equation}
\varphi^2_c(z) =
F^2_{\pi}\left[1-\left(\frac{4}{\lambda_R}+
F\left[\frac{\mu}{m},\frac{M_{\pi}}{m}\right]\right)
\frac{m^2}{F^2_{\pi}\; \cosh^2[mz]}\right]
\label{selfcons3dspino}
\end{equation}

In this case by varying $\mu \; ; \; m$ we can provide an initial
self-consistent inhomogenous state that probes field configurations
with large amplitudes away from the vacuum $\varphi_c=F_{\pi}$ in
either
direction.

 The band of unstable perpendicular wave-vectors will result in
profuse production of soft pions through the spinodal
instabilities with
a likely enhancement of pions in the low $p_T$ region.  Although these
spinodal instabilities could also be present in the one
dimensional problems
studied in the previous section, in that case there is only one mode that is
unstable. In addition, integrability forbids particle production in $ 1 + 1 $
dimensional sine-Gordon field theory.
Physically the instability of one mode  is not very interesting in the
one-dimensional models where we are interested in the dynamics of
soliton states. However it is  certainly interesting and physically motivated in
the
sigma model in three spatial dimensions where we are interested in the dynamics
of pion production from a coherent initial state. Furthermore, these
spinodal instabilities associated with the Klein paradox
phenomenon are not particular to the trial profiles studied here but
will necessarily be a robust feature of inhomogeneous states for which
large
amplitudes of the order parameter can probe the negative energy
continuum within a localized region in space, resulting in particle
production.

Fig.2 shows the profile for $\varphi_c(z)/F_{\pi}$ for a set of parameters that
allows to probe near the false vacuum near the origin, and fig. 3
shows $\varphi_c(z)/F_{\pi}$ for a different set of parameters that
probes a region with
large amplitudes away from the vacuum. Both configurations probe large
amplitude
field configurations away from the vacuum in a localized region in
space of width
$\Delta z \approx 1/m$.

\section{\bf Dynamical Evolution:  summary of renormalized update equations  
and initial conditions}
Having provided the initial state which furnishes the Cauchy data for the
proper Green's functions and addressed the renormalization issues we are now
in position to provide the update algorithm to study the dynamical evolution.
We begin by gathering all of the equations of motion and initial conditions for
$\varphi_c(\vec x,t)$ and the Green's functions which are given by eqs.
(\ref{classicaleqn},\ref{heiseneqn},\ref{firsteqn}-\ref{kkdot})
\begin{eqnarray}
&& \dot{\varphi}_c(\vec x,t) = \pi_c(\vec x,t) \label{classi} \\
&&\dot{\pi}_c(\vec x,t) = \nabla^2\varphi_c - {\cal
V}(\varphi_c,G)\label{piclassi} \\
 && \dot{G}(\vec x, \vec x';t,t)   =
D(\vec x, \vec x',t)+\tilde{D}(\vec x, \vec x',t)
 \label{firsteqn2} \\
 && \dot{D}(\vec x, \vec x',t)+\dot{\tilde{D}} (\vec x, \vec x',t) =
2K(\vec x, \vec x',t)+
\left(\nabla^2_x+\nabla^2_{x'} -M^2(\vec x,t) -M^2(\vec
x',t)\right)G(\vec x,\vec x',t) \label{ddot2} \\
&& \dot{K}(\vec x, \vec x',t) =  \left(\nabla^2_x-M^2(\vec
x,t)\right)D(\vec x,\vec x',t)+\left(\nabla^2_{x'}-M^2(\vec
x',t)\right)\tilde{D}(\vec x,\vec x',t) \label{kdot2}
\end{eqnarray}
where we have made explicit that $\cal{V}$ depends on $\varphi_c(\vec
x,t)$ and $G(\vec x,\vec x,t)$ and is given by the appropriate
expressions in the
$1+1$ dimensional cases (see eqns.  
(\ref{poten},\ref{singormass},\ref{clasfi4},\ref{heisfi4})) or by $M^2(\vec  
x,t)= M^2(\varphi_c,G)$ in the
$3+1$ dimensional case in the large $ N $ limit (see eqn.  
(\ref{flucslargen})). The initial conditions
chosen for the
dynamics of the expectation value are given by

\begin{equation}
\varphi_c(\vec x,t=0) = \varphi_{c,sc}(\vec x) \; \; ; \; \;
\pi_c(\vec x,t=0)=0
\label{initialclassi}
\end{equation}
where $\varphi_{c,sc}(\vec x)$ is the self-consistent profile given by
 eqs. (\ref{ficons}) and (\ref{cond2}) for a topological profile or
 (\ref{lumpSG}) for a lump profile in the Sine-Gordon case, or the
 profile
given in eq. (\ref{selfish}) with (\ref{alfafi4}) for the $\phi^4$ theory in
$1+1$ dimensions. In the $3+1$ dimensional sigma model in the large N,
 the profile
for the expectation value is given by eq. (\ref{selfcons3d}) in the
 stable case,
i.e. without spinodal instabilities and by
 eq. (\ref{selfcons3dspino}) in the spinodally unstable case.

In order to update the equations we need to obtain the value of the kernels
$D,\; \tilde{D}, \; K$ at the initial time $t=0$. This is achieved by using
the Heisenberg equations of motion (\ref{heiseneqn}), the field expansion
given by (\ref{psiexpansion},\ref{piexpansion}) in
terms of the eigenmodes $\psi_k$ of the differential
eq. (\ref{modeeqn})  and the initial correlations (\ref{inidistri}).
We must distinguish between the different cases without and with spinodal
instabilities.

\subsection{\bf Without spinodal instabilities}
In this case we find  the following expressions for the kernels at the
initial time:
\begin{eqnarray}
G(z,z',t=0) & = & \frac{\psi_0(z)\psi_0(z')}{2\omega_0}+ \int \frac{dk}{2\pi}
\frac{\psi_k(z)\psi^*_k(z')}{2\omega_k} \; \; \; \; \; \; \;
{\mbox{in}} \; 1+1 \; {\mbox{dimensions}} \label{gini1d} \\
G(\vec x,\vec x',t=0) & = & \int \frac{d^2k_{\bot}}{(2\pi)^2}
\frac{\psi_k(\vec x)\psi^*_k(\vec x')}{2\omega^0_k}+ \int \frac{d^3k}{(2\pi)^3}
\frac{\psi_k(\vec x)\psi^*_k(\vec x')}{2\omega_k} \; \; {\mbox{in}} \;
3+1 \; {\mbox{dimensions}} \label{gini3d}
\end{eqnarray}
 Since
the eigenmodes are rather simple in form  the integrals above can be carried
out analytically in all cases, thus providing the initial Cauchy data for the
Green's function.

The initial data for the other kernels is determined by

\begin{eqnarray}
 && D(\vec x,\vec x',t=0)  =
-\frac{i}{2}\sum_k \psi_k(\vec x) \psi^*_k(\vec x')= -
\frac{i}{2} \delta^2 (\vec x- \vec x') \label{D0} \\
 && \tilde{D}(\vec x,\vec x',t=0)  =   \frac{i}{2}\sum_k
\psi_k(\vec x) \psi^*_k(\vec x')= \frac{i}{2} \delta^2
(\vec x- \vec x') \label{tildeD0}\\
 && K(\vec x, \vec x',t=0)  =   \sum_k \frac{\omega_k}{2}
\psi_k(\vec x) \psi^*_k(\vec x')\label{K0} \\
&& \left[\nabla^2_x - M^2(\vec x,t=0)\right] G(\vec x,\vec x',t=0) =
\left[\nabla^2_{x'} - M^2(\vec x',t=0)\right] G(\vec x,\vec x',t=0) =
\nonumber \\
&& -\sum_k \frac{\omega_k}{2}\psi_k(\vec x) \psi^*_k(\vec x')\label{nablaG0}
\end{eqnarray}

We see that whereas the individual kernels evaluated at the initial
time are divergent, the combination of the kernels that enter in the
update equations above are {\em finite} and lead to the following
result for the initial value problem
\begin{eqnarray}
&&\dot{G}(\vec x,\vec x',t=0) =0 \label{dotG0} \\
&&\dot{D}(\vec x,\vec x',t=0)= \dot{\tilde{D}}(\vec x,\vec x',t=0)= 0  
\label{dotD0}\\
&&\dot{K}(\vec x,\vec x',t=0) =0 \label{dotK0}
\end{eqnarray}

\noindent{\em If} the expectation value $\varphi_c(\vec x)$ were a
solution of the time independent evolution equation for the
expectation value, we would have found an equilibrium self-consistent
solution of
the quantum back-reaction problem. However, as we pointed out before,
the self-consistent profile for $\varphi_c(\vec x)$ is {\em not} a
static solution of the evolution equation and in the second step of the
update, $\varphi_c(\vec x)$ will begin to evolve and with it the
effective mass squared $M^2(\vec x)$ for the quantum fluctuations.

\subsection{\bf With spinodal instabilities in $3+1$ dimensions:}
In this case case the Heisenberg field expansion is as given in
eq.(\ref{modexp3d}) but with the frequencies $\omega^0_k$ replaced by
the frequencies given by eq. (\ref{smallk}-\ref{Delta2}). In this
case we find the same expression for either the $1+1$ or $3+1$
dimensional cases
where the frequencies $\omega^0_k$ in the first term in eq. (\ref{gini3d}) are
given by eqs.(\ref{smallk}-\ref{Delta2}).
\begin{eqnarray}
&&\dot{G}(\vec x,\vec x',t=0) =0 \label{dotG01} \\
&&\dot{D}(\vec x,\vec x',t=0)+ \dot{\tilde{D}}(\vec x,\vec x',t=0)=
\int \frac{d^2k_{\bot}}{(2\pi)^2}\;\Theta(\Delta^2-\vec{k}^2_{\bot})\;
\frac{\mu^2+m^2-M^2_{\pi}}{\Omega^0_k}\;\psi_0(\vec x)\; \psi^*_0(\vec x')
 \label{dotD01}\\
&&\dot{K}(\vec x,\vec x',t=0) =0 \label{dotK01}\; .
\end{eqnarray}

Obviously the nonvanishing and positive expression (\ref{dotD01})
evaluated at $\vec x = \vec x'$ which corresponds to the second
derivative of $G$ with respect to time is a manifestation
of the early exponential growth of the quantum fluctuations associated
with the spinodal instabilities and particle production.

The systems of update equations  shown above displays explicitly that
although the individual kernels $K, D$ have short distance singularities, these
cancel in the combinations that are necessary for the update equations.
Insofar as the Green's function $G(\vec x,\vec x',t)$ is evaluated at separate
spatial points it remains finite throughout the evolution since the
initial value
is finite and the update equations are in terms of finite
quantities. However, once the coincidence limit is taken, quadratic
and logarithmic divergences will
appear in the short distance operator product expansion. From the time evolved
value of $G(\vec x,\vec x',t)$ we must now take the coincidence limit
to construct
the updated effective mass $M^2(\vec x,t)$ and the short distance
divergences must
be renormalized.
However our previous analysis of the short distance divergences
determines unambiguously the manner to obtain the fully renormalized equations.
In $1+1$ dimensions the equations of motion are rendered finite by
mass renormalization and the replacement
\begin{equation}
G(z,z) \rightarrow = G_{sub}(z,z)= G(z,z)-\int^{\Lambda}_0 \frac{dk}{2\pi}
\frac{1}{\sqrt{k^2+M^2}}\end{equation}
where $\Lambda$ is the upper momentum cutoff. In three dimensions we can take
the spatial coincidence limit of the Green's function in the {\em finite}
quantity $M^2_{\pi}+ \delta M^2(\vec x,t)$ with $\delta M^2(\vec x,t)$ given by
the expression (\ref{finitemass}).

\section{Numerical Strategy:}
Although implementing a full numerical study of the evolution equations with
the self-consistent initial data is beyond the scope of this paper, we
briefly describe how we will implement numerically the evolution in a
forthcoming study.
The implementation is obviously much simpler in the one dimensional case: one
begins by discretizing space introducing a lattice cutoff $a$ and the
upper momentum cutoff becomes $\Lambda = \pi/a$, so that the integral
for $G_0$ can
now be done at once leading to function of $Ma$. Since eventually we
are interested in the coincidence limit of $G(z,z')$ we do not need to
update the
functions of two variables $z,z'$ for the whole range of the two variables, the
non-locality of the update equations are mild only requiring the derivatives
at separate points. Thus for fixed $z$ we can simply update the equations for
$z'$ just a few lattice spacings away in such a way that the
discretized derivative does not introduce coincidences. The
coincidence limit will correspond
to writing $z' \rightarrow z+a$ and the renormalized Green's function
is obtained
by subtracting the value of $G_0$, in the limit when $Ma <<1$ the renormalized
Green's function is insensitive to the short distance cutoff.

In $3+1$ dimensions the strategy consists of the discretization of
space in a cubic lattice of equal spacing $a$ along all the
directions. Now the second
derivatives needed in the update equations require storing the field
configurations on 6 sites around the particular points $\vec x, \vec
x'$, again since ultimately we are interested in local field
configurations and the spatial
coincidence limit we do not need to store the configurations for all
of the range
of $\vec x'$ but only in the immediate neighborhood of the point $\vec
x$. For a
lattice spacing $a$ there corresponds an upper momemtum
$\vec{k}_{max}= (\pi / a,\pi / a,\pi / a)$ and the momentum integrals
needed for $G_0$ are done
with spherical symmetry and a radial  upper momentum cutoff
$\Lambda = \sqrt{3} \pi / a$ which is the radius of the spherical
Brillouin zone. The
coincidence limit must be understood now as $G(\vec x, \vec x + \vec
a)$ and for
$Ma << 1$ the renormalized quantity $M^2(\vec x,t)$ is insensitive to the
short distance cutoff.

As discussed in the previous section within the context of
renormalization, the linear sigma model is a  low energy theory with a
cutoff $\Lambda \leq 2-3 \;\mbox{Gev}$ and this restricts the value of
the lattice spacing for a numerical analysis. There is, however, a
much more stringent constraint on the sigma model description of low
energy physics: this model does not include other important hadronic
degrees of freedom such as the pseudoscalars $\eta \; ; \eta'$ with masses  
$\approx 500-600\;  \mbox{Mev}$ and vector mesons (which determine
pion electromagnetic form factors) with masses $\approx 700-800 \;  
\mbox{Mev}$. These
constraints suggest that the actual physical cutoff for the sigma
model is not much bigger than $\Lambda \approx 500-700 \; \mbox{Mev}$
resulting in a lattice spacing such that $M_{\pi}a \approx 0.2$. Thus
the combination of Landau pole arguments and low energy constraints on
the
dynamics require that a physically reasonable lattice spacing for a
numerical treatment be of the order of  $\approx 0.2 \; \mbox{fm}$,
which will not result in strong short distance effects in the update
of the dynamical equations.

\section{Conclusions:}

In this article we focused on establishing a consistent framework to study
the non-equilibrium evolution of inhomogeneous non-perturbative field
configurations including quantum backreaction effects. The main goal
is to provide numerically implementable schemes to study the time
evolution of large amplitude field configurations including the
quantum backreaction associated  with particle production on the
relaxational dynamics.

We proposed a self-consistent variational approach to study the
non-perturbative aspects of the evolution of large amplitude
inhomogeneous configuration, and analyzed the available frameworks to
implement them. Since our main goal is to provide a {\em practical}
method amenable to numerical implementation we highlighted in detail
the advantages and shortcomings of the different approaches. We
recognized the main ingredients that are required for a practical
numerical implementation of this program: i) a set of {\em local}
dynamical update equations both in space and time,  ii) a
self-consistent solution of the initial value problem that provides
the Cauchy data at the initial
time, iii) renormalizability
implemented at the level of the update equations, so as to ensure insensitivity
to the lattice introduced in the numerical implementation. After
analyzing several
different manners to implement a variational, self-consistent treatment of the
non-equilibrium dynamics for these inhomogeneous problems, we
identified a formulation that leads to a closed set of {\em local}
update equations. We then studied in
detail the renormalizability of this set of equations in both one and
three spatial dimensions.  The study of the dynamics of
non-perturbative inhomogeneous field configurations  in one spatial
dimension is warranted by the interest in understanding the dynamics
of soliton configurations in the
presence of quantum excitations, but also as model systems to test
these methods in simpler settings. In one spatial dimension   we
provide simple  self-consistent ansatz  for the initial
Cauchy data in terms of exactly solvable
potentials for the Sine-Gordon and $\phi^4$ model. In the Sine Gordon
case the exact mass spectrum, S-matrix, form factors are
known and this allowed us to compare this simple ansatz
to  exact quantum results.  Numerical resolution of the full set of
variational equations should provide better approximations.

In three dimensions we focused on the linear sigma model in the large
$ N $ limit as a model for relaxation of non-perturbative coherent pion
configurations such as
disoriented chiral condensates. In this case we provided a
self-consistent solution for the initial Cauchy data in the case of
cylindrical symmetry by extending the results from the one dimensional
examples.

We found that for large amplitude configurations a mechanism akin to
the Klein paradox resulting in particle production is responsible for
novel spinodal instabilities that will lead to profuse pion production
at low $p_t$ through the relaxation of the inhomogeneous condensate.

 Furthermore we provided the full set of renormalized (semi)local
 update equations  that can be implemented numerically and discussed
 in detail the strategy for a practical implementation, which is the
 next step of the program and currently in progress.

\acknowledgements
D. B. thanks the N.S.F for partial support through the grant
awards: PHY-9605186 and  LPTHE for warm hospitality.
F. C thanks the DOE for support, P. S. thanks the INFN for support.
F.C. would like to thank Emil Mottola and Salman Habib for several useful
discussions and A. Kerman for valuable conversations.

D. B. and H. J. de V.  acknowledge partial support from NATO.




\begin{center}
\begin{figure}[t]
 \epsfig{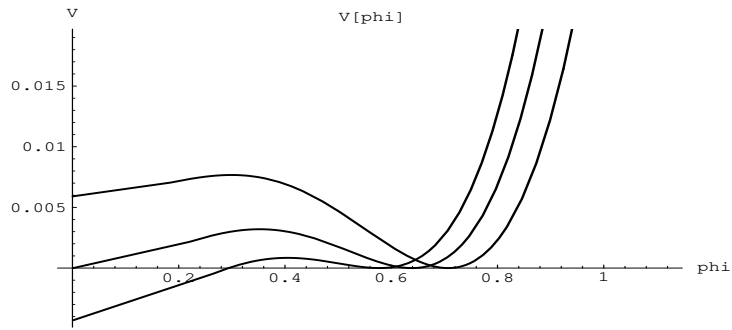}
\caption{Variational effective potential for $\phi^4$ in $1+1$ dimensions for
$g/m^2_r=1,\; g_c,\; 2$} 
\end{figure}
\end{center}

\begin{center}
\begin{figure}[t]
 \epsfig{file=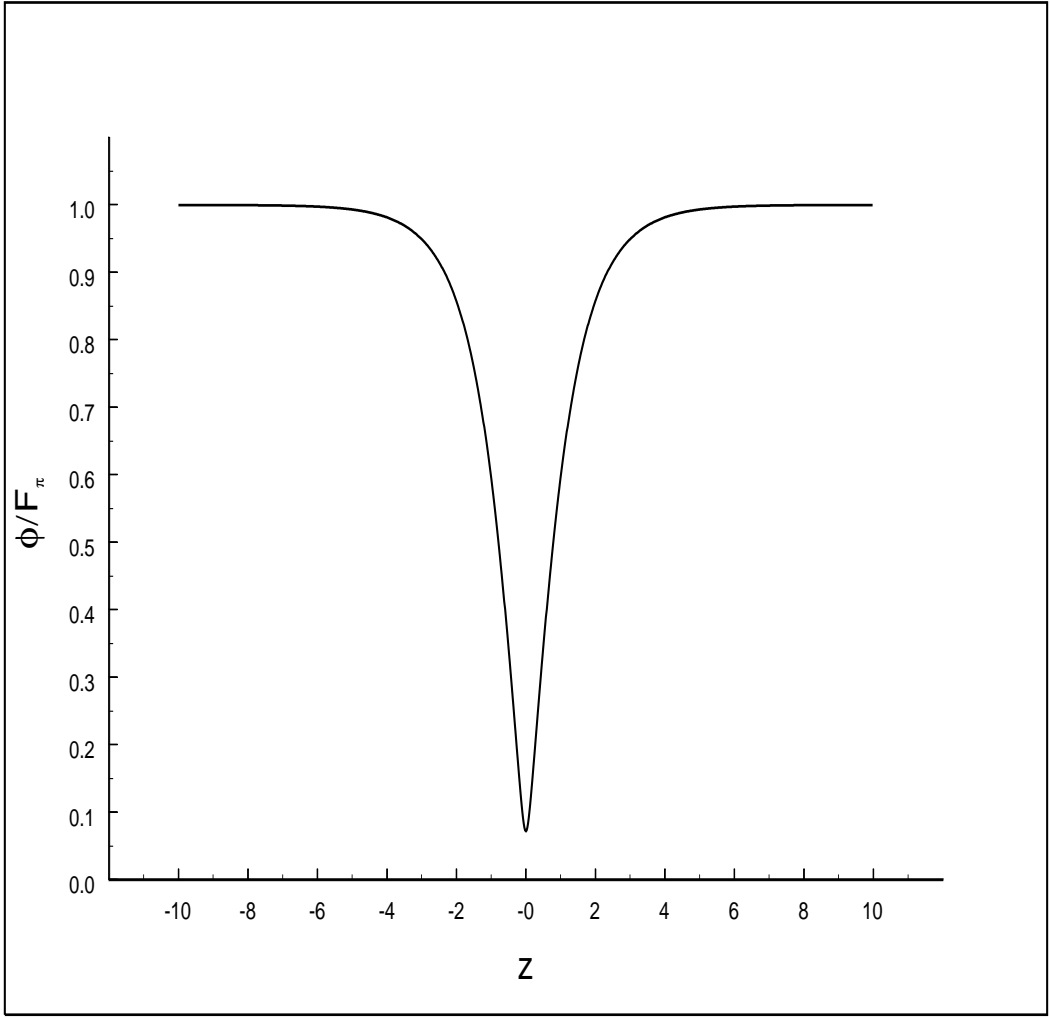,width=6in,height=6in}
\caption{$\varphi_c(z)/F_{\pi}$ vs. $mz$ for $M_{\pi}/m=0.575,\; \mu=0\; \lambda_R=36$} 
\end{figure}
\end{center}

\begin{center}
\begin{figure}[t]
 \epsfig{file=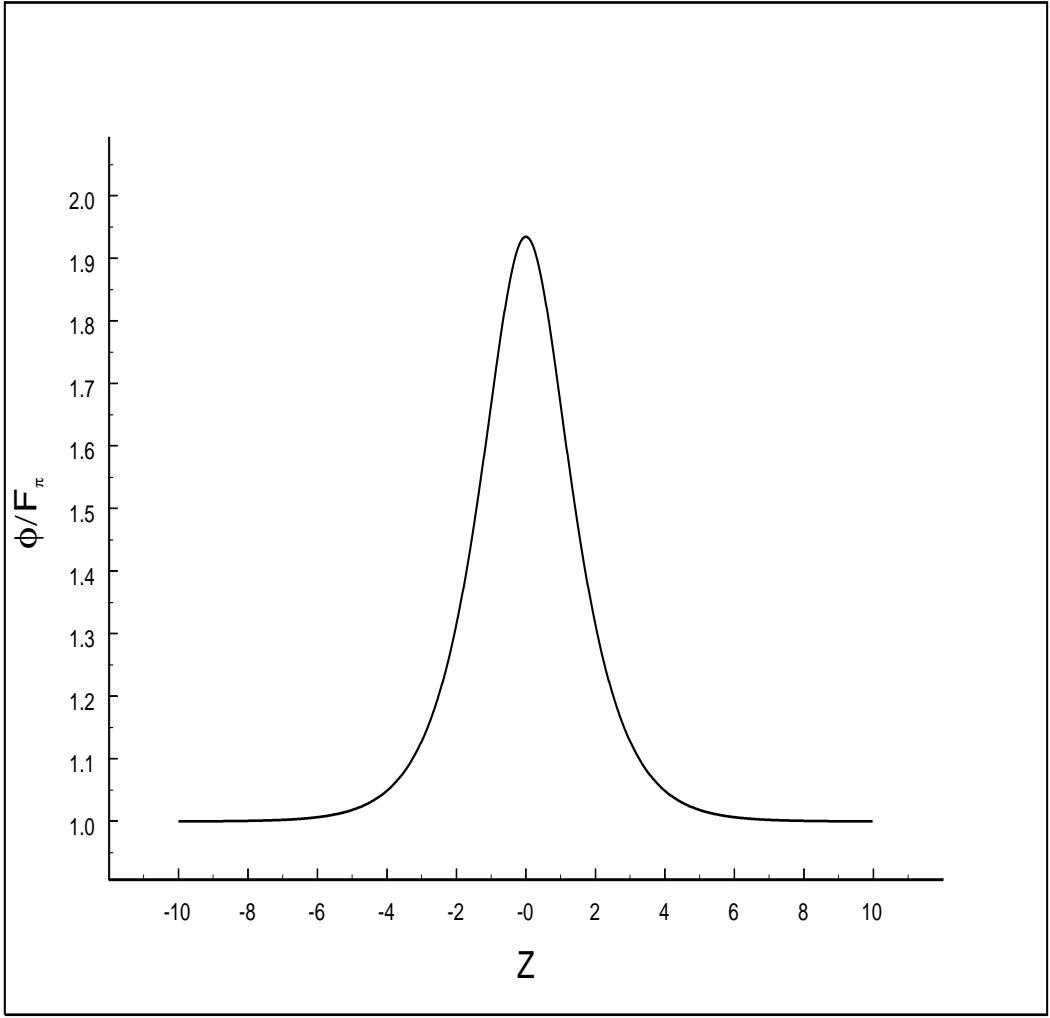,width=6in,height=6in}
\caption{$\varphi_c(z)/F_{\pi}$ vs. $mz$ for $M_{\pi}/m=0.0019,\; \mu=0 \; ,\lambda_R=36$} 
\end{figure}
\end{center}

 
\end{document}